%% file: main.tex
\documentclass[fleqn,usenatbib]{mnras}

\usepackage{newtxtext,newtxmath}

\usepackage[T1]{fontenc}
\usepackage{ae,aecompl}

\usepackage{graphicx}	
\usepackage{amsmath}	
\usepackage[usenames,dvipsnames]{xcolor}
\usepackage{hyperref}
\usepackage[normalem]{ulem}
\usepackage{soul}

\newcommand{\sub}[1]{_{\mathrm{#1}}}
\newcommand{\msun}{M\sub{\sun}}
\newcommand{\nbody}{$N$-body }
\newcommand{\hiz}{high-$z$ }

\def\equationautorefname~#1\null{Eq.~(#1)\null}
\def\figureautorefname~#1\null{Fig.~#1\null}
\newcommand{\appref}[1]{\hyperref[#1]{Appendix~\ref{#1}}}

\title[Dark matter core formation by baryonic clumps]{Dark matter cores in massive \hiz galaxies formed by baryonic clumps}

\author[G. Ogiya \& D. Nagai]{
Go Ogiya$^{1,2,3}$\thanks{E-mail: \url{gogiya@zju.edu.cn} (GO)}
and 
Daisuke Nagai$^{4}$
\\
$^{1}$Waterloo Centre for Astrophysics, University of Waterloo, Waterloo, ON N2L 3G1, Canada \\
$^{2}$Department of Physics and Astronomy, University of Waterloo, 200 University Avenue West, Waterloo, Ontario N2L 3G1, Canada \\
$^{3}$Institute for Astronomy, School of Physics, Zhejiang University, Hangzhou 310027, China \\
$^{4}$Department of Physics, Yale University, New Haven, CT 06520, U.S.A.
}

\date{Accepted XXX. Received YYY; in original form ZZZ}

\pubyear{2022}

\begin{document}
\label{firstpage}
\pagerange{\pageref{firstpage}--\pageref{lastpage}}
\maketitle

\begin{abstract}
The rotation curves of some star forming massive galaxies at redshift two decline over the radial range of a few times the effective radius, indicating a significant deficit of dark matter (DM) mass in the galaxy centre. The DM mass deficit is interpreted as the existence of a DM density core rather than the cuspy structure predicted by the standard cosmological model. A recent study proposed that a galaxy merger, in which the smaller satellite galaxy is significantly compacted by dissipative contraction of the galactic gas, can heat the centre of the host galaxy and help make a large DM core. By using an \nbody simulation, we find that a large amount of DM mass is imported to the centre by the merging satellite, making this scenario an unlikely solution for the DM mass deficit. In this work, we consider giant baryonic clumps in high redshift galaxies as alternative heating source for creating the baryon dominated galaxies with a DM core. Due to dynamical friction, the orbit of clumps decays in a few Gyr and the baryons condensate at the galactic centre. As a back-reaction, the halo centre is heated up and the density cusp is flattened out. The combination of the baryon condensation and core formation makes the galaxy baryon dominated in the central 2-5 kpc, comparable to the effective radius of the observed galaxies. Thus, the dynamical heating by giant baryonic clumps is a viable mechanism for explaining the observed dearth of DM in high redshift galaxies. 
\end{abstract}





\section{Introduction}
\label{sec:intro}

Observations in the last decades have revealed interesting properties of high redshift (high-$z$) star forming galaxies, including the domination of baryon mass over dark matter (DM) mass and the deficit of DM mass, in contrast local counterparts. Based on the observed gas density field and the rotation curve of emission lines, such as H$\alpha$ and CO, it has been inferred that massive star forming galaxies in the epoch of the cosmic noon (i.e., $z \sim 2$) are dynamically dominated by baryons \citep[][and references therein]{Genzel2006,vanDokkum2015,Genzel2017,Ubler2018,Genzel2020,Price2021}. They are typically rotation supported disk galaxies with $V\sub{rot} / \sigma = 2-6$ \citep{ForsterSchreiber2018,Price2020}, where $V\sub{rot}$ and $\sigma$ represent the rotation velocity and the velocity dispersion of the galactic gas, and the rotation velocity spans a range of $V\sub{rot} = 100-350$~km/s. Similar to the \hiz star forming galaxies, passive galaxies at high- \citep{Mendel2020} and low redshift \citep{Tortora2022} are baryon dominated, while the baryon dominance of local disk galaxies is less significant. Star forming galaxies at $z \sim 1$ appear to be in the transitional state \citep{Sharma2021}. 

The rotation curve of some individual star forming \hiz galaxies declines at galactocentric radii larger than a few times the effective radius \citep{Genzel2017,Genzel2020,Price2021}, indicating a significant deficit of DM mass in the galaxy centre. Similar declining rotation curves are obtained by using a stacking approach (\citealt{Lang2017}, but see also \citealt{Tiley2019}). Even when considering the pressure force of the non-circular motion (e.g., asymmetric drift) in the galactic gas \citep{Burkert2010,Burkert2016}, the gap between the theoretically predicted DM mass and the observationally inferred mass is significant. The surrounding halo, lacking the central DM mass of $\sim 30$ percent of the bulge mass, is interpreted to have a constant mass density core \citep{Genzel2020}.

A similar DM mass deficit problem that has been discussed for nearby dwarf galaxies could provide a hint to understanding the origin of baryon dominated \hiz galaxies. Cosmological \nbody simulations based on the $\Lambda$ cold dark matter ($\Lambda$CDM) model predict a divergent mass density structure, the so-called cusp, at the centre of DM haloes \citep[e.g.,][]{Navarro1997,Springel2005_millennium,Stadel2009} in a broad halo mass scale and in a wide redshift range \citep[][and references therein]{Wang2020,Ishiyama2021}. However, observations have indicated that the DM mass in the centre of nearby dwarf galaxies is smaller than the theoretical expectation \citep[e.g.,][]{Flores1994,Burkert1995,Swaters2003}, while rotation curves have a diversity \citep[][and references therein]{Oh2015,Oman2015}.

A potential solution to resolve the mismatch between the $\Lambda$CDM prediction and observations is the dynamical heating caused by the potential fluctuation of the galactic gas driven by supernova feedback and radiative cooling \citep[e.g.,][]{Pontzen2012,Ogiya2014,Read2016}. Observations showed that star formation induced cores can be formed in the centre of galaxies at $z \sim 1$ \citep{Bouche2021}. However, supernova feedback would not be powerful enough to expel the gas from the centre of massive galaxies, and thus this process does not work to alter the central DM density structure of massive systems \citep{Silk2012,DiCintio2014}. Although the baryon dominated \hiz galaxies have been seen in cosmological hydrodynamical simulations \citep{Lovell2018,Teklu2018,Ubler2021}, the physical origin of such systems remains unclear.

Idealised simulations systematically investigating the impact of specific physical processes at a time can shed light on the essential processes to originate baryon dominated \hiz galaxies. By using a semi-analytical model of \cite{Jiang2021} and idealised \nbody simulations, \cite{Dekel2021} considered a combination of a dry galaxy merger and feedback from active galactic nuclei (AGN; see also e.g., \citealt{Peirani2017}) as a heating source to reduce the central DM mass and argued that this combination of processes may transform a central density cusp to a core with a size of $\sim 10$~kpc. In their model, both the larger host and smaller satellite galaxies have a central cusp prior to the merger, and the satellite is supposed to be a compact system, like a blue nugget galaxy, with a high density, making it resilient to the tidal force of the host. This condition enables the satellite to survive until it closely encounters and dynamically heats up the central cusp of the host. The dynamical heating of galaxy mergers makes the central DM particles more vulnerable to the additional heating induced by AGN feedback and helps to make a large DM core. If the satellite has a lower density as predicted by the concentration-mass-redshift relation \citep[e.g.,][]{Diemer2015,Ludlow2016}, the heating efficiency of the galaxy merger is significantly lowered and the central cusp remains even if AGN feedback is considered. Although dry mergers can heat the cusp of the host, the merging satellite supplies DM mass to the centre of the merger remnant, especially when the satellite mass is not reduced by the tidal interaction with the host. This can hamper the resultant reduction of the central DM density of the merger remnant. The theoretical framework by \cite{Dehnen2005} predicts that dry mergers do not produce a merger remnant having a central density lower than those of progenitors, and it has been confirmed by \nbody simulations  \citep{Boylan-Kolchin2004,Kazantzidis2006,Ogiya2016b,Angulo2017,Drakos2019}. 

In this paper, we consider another possible heating source in \hiz star forming galaxies, giant baryonic clumps. More than half of \hiz star forming galaxies have a disky morphology with irregular substructures (clumps), and such galaxies are often referred to as clump clusters or chain galaxies \citep[e.g.,][]{Cowie1995,vandenBergh1996,Moustakas2004,Elmegreen2005,Elmegreen2006}. The clumps are massive ($[10^8:10^9]~\msun$) and compact \citep[$\la 1$~kpc; e.g.,][]{Elmegreen2009,ForsterSchreiber2011,Genzel2011}, and thus they are dense enough to survive in the strong tidal field of the galactic centre. The orbital energy and angular momentum of massive objects orbiting in a larger system, including the baryonic clumps, will be transferred to material of the larger system (host galaxy) through dynamical friction \citep{Chandrasekhar1943}, which in turn can heat and flatten the central density cusp \citep{El-Zant2001,El-Zant2004,Goerdt2010,Inoue2011}. 

The primary goal of this paper is investigating how the density structure of the host is altered by the dynamical heating of giant baryonic clumps and how the mass composition at the galactic centre evolves. To this end, we use a series of \nbody simulations (clump simulations) in which giant clumps orbit within a host galaxy. Since giant clumps consist of baryons, their condensation leads to the baryon domination at the galactic centre. We also examine the {\it total} density structure of the remnant of the galaxy merger, which was not studied by \cite{Dekel2021}. We use another \nbody simulation (merger simulation) in which a larger host galaxy interacts with a satellite galaxy. While the dynamical impact of a galaxy merger can reduce the central density of the group of particles belonging to the larger host galaxy prior to the merger, the resultant central DM density increases by the DM mass supply from the dense satellite.

The rest of the paper is organised as follows. \autoref{sec:sim_basics} describes the basic elements of our $N$-body simulation models. In \autoref{sec:merger_sim}, we study the {\it total} density profile of the system experienced a dry merger. We discuss the role of baryonic giant clumps in forming a baryon dominated galaxy in \autoref{sec:clump_sim}. The results are summarised in \autoref{sec:summary}. Throughout the paper, the cosmological parameter set obtained by \cite{Planck2016} is adopted.

\section{Simulation Basics}
\label{sec:sim_basics}

In this Section, we describe the common basic elements of our \nbody simulations. All systems considered in the simulations (host and satellite galaxies in the merger simulation or host galaxy and giant clumps in the clump simulations) are supposed to follow initially the Navarro-Frenk-White density profile which has a central cusp \citep[][hereafter NFW]{Navarro1997},
\begin{equation}
    \rho(r) = \frac{\rho\sub{s}}{(r/r\sub{s})(1+r/r\sub{s})^2},
        \label{eq:nfw_rho}
\end{equation}
where $r$, $\rho\sub{s}$ and $r\sub{s}$ are the distance from the centre, the scale density and the scale length of the density distribution, respectively. As indicated, the NFW density profile is a two-parameter model and we employ the concentration and the mass of the system as the structural parameters. The concentration of the system is given by $c\equiv r\sub{200}/r\sub{s}$, where $r\sub{200}$ is the virial radius whose enclosed mean density is 200 times the critical density of the universe, $\rho\sub{crit}(z)$ at redshift $z$, and the mass enclosed within $r\sub{200}$ is the virial mass, 
\begin{equation}
    M\sub{200} \equiv \frac{800 \pi}{3} \rho\sub{crit}(z) r\sub{200}^3.
        \label{eq:vir_mass}
\end{equation}

We draw the initial position and velocity vectors of \nbody particles with respect to those of the centre of the system by using the acceptance-rejection sampling method \citep{Press2002}, in which the acceptance or rejection of a quantity having a value of $x$ is determined based on a probability, $p(x)$. The distance from the centre of the system to a particle, $r$, is stochastically sampled in the radial range of $r \leq r\sub{200}$, based on $p(r) \propto \rho(r)r^2$. Then, we randomly draw a unit vector to specify the position vector of the particle. The energy of each particle, $e$, is stochastically sampled with the phase-space distribution function, $f(e)$, numerically computed using the Eddington formula \citep{Eddington1916}, i.e., $p(e, r) \propto f(e)|e-\Phi(r)|^{1/2}$ \citep{Binney2008}, where $\Phi(r)$ is the gravitational potential profile of the system. Since the phase-space distribution function is assumed to depend only on energy, the direction of the velocity vector of each particle is randomly drawn. 

In all simulations, the virial mass of the host system is $M\sub{200,host} = 3 \times 10^{12} \msun$ which yields $r\sub{200,host} = 146$~kpc at $z = 2$ and its concentration is $c\sub{host} = 5$. Note that smooth baryonic components, such as the central bulge and stellar and gas disks, of the host system are approximated as part of the NFW halo. The setup is the same as those employed in \cite{Dekel2021}. In this study, we employ the approximated model for simplicity and leave the inclusion of the baryonic components for future studies. The host system is modelled with 67,108,864 particles, and each particle has a mass of $m\sub{p} \approx 4.5 \times 10^4 \msun$. Employing the large number of particles, we have the convergence radius of $r\sub{conv} = 0.3$~kpc \citep{Power2003}. Here, $r\sub{conv}$ is defined as the radius at which the simulation time of $\sim 3.3$~Gyr, corresponding to the age of the universe at $z=2$, is equal to the two-body relaxation time,
\begin{equation}
    T\sub{rel}(r) = \frac{\pi N(r)}{4 \ln{N(r)}} \biggl [ \frac{r^3}{GM(r)} \biggr ]^{1/2}, 
        \label{eq:trel}
\end{equation}
where $N(r)$ and $M(r)$ are the number of particles belonging to the host and the host mass contained within $r$, respectively. Since the two-body relaxation time monotonically increases with radius, the nature of collisionless dynamics is guaranteed at $r \geq r\sub{conv}$. An artificial density core can be formed due to two-body relaxation at $r < r\sub{conv}$.

For \nbody computation, we use a code employing the oct-tree algorithm \citep{Barnes1986}. The gravity computation is accelerated with Graphics Processing Units \citep{Ogiya2013}. In the simulations, a Plummer force softening \citep{Plummer1911} of $\epsilon=0.02$~kpc is employed. Simulations varying the number of particles, equivalently the mass resolution, or the force softening reveal that our simulation results are numerically converged. We employ the cell opening criteria by \cite{Springel2005_gadget2} with a parameter controlling the force accuracy of $\alpha=0.01$. The timestep is updated in each iteration by following the prescription of \cite{Power2003} and shared with all particles. We use the second order Leapfrog scheme for the orbit integration of particles. The centre and bulk velocity of the \nbody systems are tracked with the method outlined in \cite{vandenBosch2018}, while re-binding of particles is not allowed.\footnote{The scheme in \cite{vandenBosch2018} allows particles having a positive binding energy at the previous snapshot to be re-bound if they have a negative binding energy at the present snapshot. Numerical experiments imply that \cite{Dekel2021} may not allow re-binding of particles.}

\section{Re-examining the impacts of mergers}
\label{sec:merger_sim}

\subsection{Set up of the merger simulation}
\label{ssec:merger_setup}

The merger simulation imitates the simulation of a merger event between two systems performed by \cite{Dekel2021}, and re-examines the impact of dry mergers onto the density profile of the merger remnant. The parameters for the host system are described in \autoref{sec:sim_basics}. In this simulation model, we neglect the bulge and stellar and gas disks of the host. If those components enhance the central density of the host system, the DM density cusp would be more resilient to the dynamical heating by the merger. The satellite mass is $M\sub{200,sub} = M\sub{200,host}/8 = 3.75 \times 10^{11} \msun$, which is slightly larger than what \cite{Dekel2021} employed ($M\sub{200,sub} = M\sub{200,host}/10$), and the virial radius is $r\sub{200,sub} = 73$~kpc. The concentration of the satellite is $c\sub{sub}=50$ modelling the compaction effect, driven by dissipative contraction of the galactic gas \citep{Zolotov2015,Tacchella2016a}. Each particle belonging to the satellite has the same mass resolution, $m\sub{p}$, as that of host particles, and 8,388,608 particles are employed to model the satellite. 

The host-centric frame in which the host is located at the origin with zero bulk velocity is considered to set the initial relative position and velocity vectors between the two systems. The satellite is initially located at the apocentre of the relative orbit. We employ a pair of dimensionless parameters to specify the merger orbit. The first parameter, $x\sub{c} \equiv r\sub{c}(E)/r\sub{200,host}$, describes the orbital energy. Here, $r\sub{c}(E)$ is the radius of a circular orbit having an orbital energy of $E$. The second parameter, $\eta \equiv L/L\sub{c}(E)$, controls the angular momentum of the merger orbit. $L$ and $L\sub{c}(E)$ are the actual angular momentum and the angular momentum of the circular orbit of $r\sub{c}(E)$, respectively. In the merger simulation, we set $x\sub{c} = 1.0$ and $\eta = 0.1$, yielding the same merger orbit as the \nbody simulation in \cite{Dekel2021}.

\subsection{Results}
\label{ssec:merger_results}

\begin{figure}
    \begin{center}
        \includegraphics[width=0.4\textwidth]{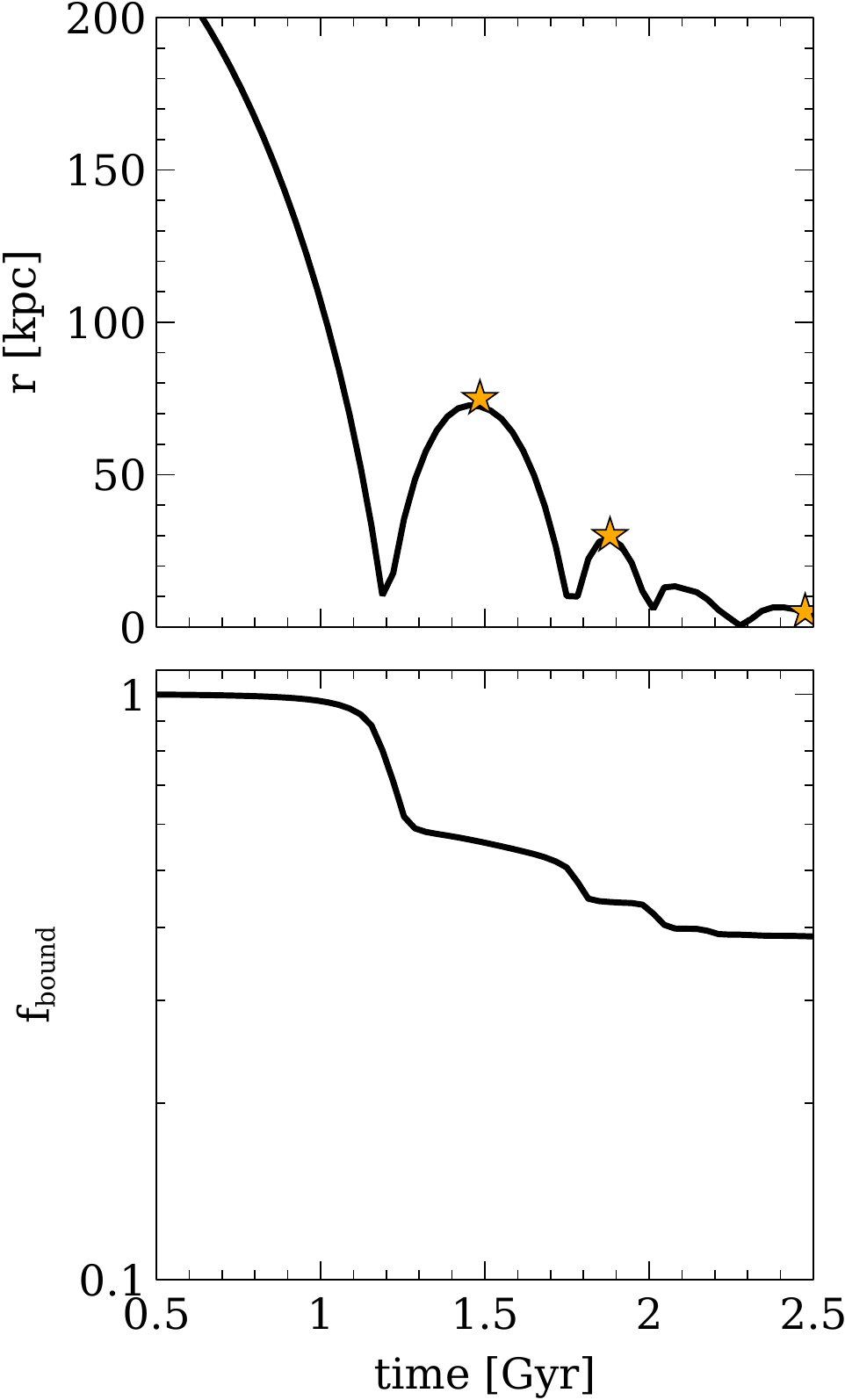}
    \end{center}
        \caption{
        {\it Upper panel:} Distance between the centres of the two systems. Stars indicate the snapshots shown in \autoref{fig:d21_rho}. The satellite orbit decays in a few orbits due to dynamical friction. 
        {\it Lower panel:} Bound mass evolution of the satellite. The tidal mass loss is almost halted after the third percentric passage ($t \ga 2$~Gyr) and $\sim 40$ percent of the initial mass is still retained in the satellite, because the compacted satellite is resilient to the tidal force. 
        \label{fig:d21_orb_mass}}
\end{figure}

First, we show the orbital evolution of the satellite in the host-centric frame in the upper panel of \autoref{fig:d21_orb_mass}. Because the host-to-satellite mass ratio is small (eight), dynamical friction works strongly to decay the satellite orbit. Consequently, the two systems are merged into a single system in a few Gyr. The lower panel presents the bound mass fraction of the satellite. The satellite mass is reduced at each pericentric passage where the satellite feels the strong tidal force of the host. However, the tidal mass loss is almost halted after the third pericentric passage ($t \ga 2$~Gyr), because the satellite has a high concentration to model the compaction effect and is denser than the host. Note that the orbital and mass evolution shown in \autoref{fig:d21_orb_mass} well reproduce those in \cite{Dekel2021}. 

\begin{figure}
    \begin{center}
        \includegraphics[width=0.4\textwidth]{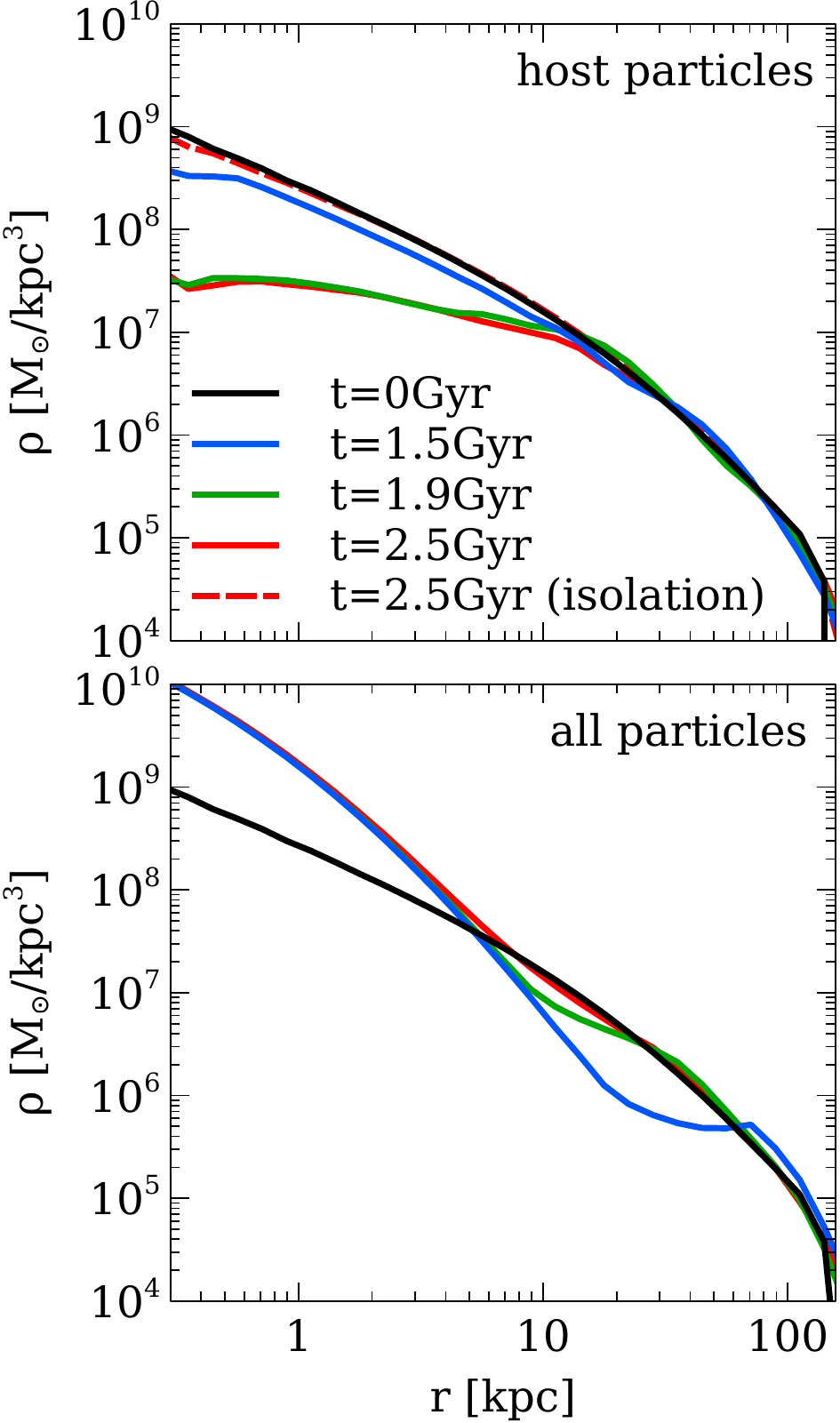}
    \end{center}
        \caption{
        Density profiles obtained from the merger simulation. In the analysis of the upper (lower) panel, particles initially belonging to the host (all particles) are included. Lines with different colours present the different phases of the dynamical evolution, as presented in the legend. To test the stability of the \nbody system, the density profile after evolution of 2.5~Gyr in isolation is shown in the upper panel (dashed red). The central density of the host is reduced by the energy and angular momentum transfer from the merging satellite and the central cusp gets shallower ({\it upper panel}). However, the total central density is in fact increased and the slope of the total density profile gets steeper at $r < 10$~kpc through the merger because the satellite having the higher density settles at the centre of the merger remnant ({\it lower panel}). 
        \label{fig:d21_rho}}
\end{figure}

Next, we demonstrate the density profile obtained from the merger simulation in \autoref{fig:d21_rho}. In the analysis of the upper panel, particles initially belonging to the host (host particles) are taken into account, i.e., particles initially belonging to the satellite are excluded from determining the centre of the system and from deriving the density profile. This corresponds to the analysis in \cite{Dekel2021}. The central cusp survives after the first close encounter between the two systems (blue), while the central density starts decreasing from the initial configuration (black), due to the re-distribution of energy and angular momentum driven by the merging satellite. In the subsequent evolution, the dynamical heating further reduces the central density of the host and the density slope gets shallower (green and solid red). In the end, the host density at $r \la 10$~kpc has been reduced, consistent with \cite{Dekel2021} as well as the prediction by an analytical model presented in \appref{app:ana_model}. Note that the host keeps its initial configuration for a long time when it is isolated (dashed red), ensuring that the \nbody host model is generated in a dynamically stable state. 

However, the dense satellite retains a large fraction of its mass and sinks to the centre of the merger remnant (\autoref{fig:d21_orb_mass}), and a large amount of DM should be, in fact, imported into the centre of the merger remnant. Therefore, we need to consider not only host particles, but also particles initially belonging to the satellite. The lower panel of \autoref{fig:d21_rho} repeats the analysis in the upper panel with all \nbody particles in the simulation. The initial profile (black) is almost identical to that shown in the upper panel, because a large fraction of satellite particles are initially located outside the virial radius of the host. After the first encounter at $t=1.5$~Gyr (blue), the centre of the merger remnant is occupied by satellite particles while the two systems are not completely merged yet, as indicated by a density bump at $r \sim 20-100$~kpc. Then the two systems are merged and the density profile gets smoother at $t=1.9$~Gyr (green) and $t=2.5$~Gyr (red). While the density structure in the outskirt returns to the initial configuration, the central density profiles stay intact after $t=1.5$~Gyr, rendering the centre of the system to be dominated by the particles associated with the infalling satellite and causing the central density to remain higher after the merger. 

We find that the expected baryon mass of the satellite, $(\Omega\sub{b}/\Omega\sub{m})M\sub{200,sub} \sim 6 \times 10^{10} ~ \msun$ is enclosed within $r \sim 3$~kpc from the centre of the merger remnant, where $\Omega\sub{m}$ and $\Omega\sub{b}$ are the density parameters for the total matter and baryon, respectively. Measuring the enclosed mass at $r = 10$~kpc, the total mass of the merger remnant at $t=2.5$~Gyr exceeds the initial mass consisting of only host particles by $\sim$ 50 percent, while the mass enhancement approaches to 12.5 percent at larger radii, as set in the configuration of the simulation. Thus, dry mergers can enhance the DM mass at $3 \la r/{\rm kpc} \la 10$ of the merger remnant even if supposing that the central part of the satellite is totally baryon dominated. In reality, the DM distribution in the satellite is concentrated as well by the strong gravity of the compacted baryon component, and the dry merger can increase DM mass at the centre of the merger remnant. 

The results from the merger simulation suggest a problem with the core formation scenario proposed by \cite{Dekel2021}. Galaxy mergers heat the density cusp and make DM particles in the host centre vulnerable to the additional heating induced by AGN feedback. The key to form a large core with this scenario is the dissipational compaction effect that works for galaxies having a dynamical mass of $\ga 10^{11.3}~\msun$ \citep{Zolotov2015,Tacchella2016a}. Such a massive satellite will sink to the centre of the merger remnant in a short time due to dynamical friction and supply a large amount of DM mass to the centre of the merger remnant. Therefore, in this scenario, the central DM cusp is removed and a DM core can form temporarily, although the central DM density is replenished by the satellite then. The timescale of the core persistence depends on the merger orbit. For this scenario to work, some fraction of \hiz galaxies with a DM core must be accompanied with their dense satellite galaxies.

\section{Role of giant clumps in forming baryon dominated galaxies}
\label{sec:clump_sim}

\subsection{Set up of the clump simulations}
\label{ssec:clump_setup}

Giant clumps are thought to be formed through fragmentation of the galactic disk \citep[e.g.,][]{Noguchi1999,Bournaud2007,Ceverino2010,Tamburello2015}, as observations revealed that the gas disks of \hiz galaxies are gravitationally unstable \citep{Genzel2014}. The formation scenario for giant baryonic clumps has been investigated extensively in the literature, with highlights on the criteria of disk fragmentation \citep{Toomre1964,Behrendt2015,Inoue2016}, the evolution of fragmented disks \citep{Krumholz2010_disk_dyn}, and the survivability of clumps \citep{Krumholz2010_clump_suv,Ginzburg2021,Dekel2022_vdi}. Hydrodynamical simulations of high resolutions showed that kpc-sized giant clumps may have internal substructures \citep{Behrendt2016,Tamburello2017}, and such smaller clumps have been observed \citep{Livermore2015,Mestric2022}. 

Giant clumps are also a key to understand the formation of rotating bulges in disk galaxies and thick disks \citep{Noguchi1999,Ceverino2010,Inoue2011,Inoue2014,vanDonkelaar2021}. Dynamical friction drives the orbital decay of giant clumps and they can transform into the bulge and thick disk. As a back-reaction, the DM density cusp is dynamically heated up, making these giant clumps a possible heating source to flatten out the cusp of the DM halo surrounding \hiz galaxies. 

Using clump simulations, we study how the central density structure of the DM host halo evolves under the existence of giant clumps. A large mass fraction of giant clumps is occupied by stars \citep{Tamburello2017}, making the \nbody technique (neglecting hydrodynamics) a reasonable approximation for modelling the dynamical evolution of DM host halo containing giant clumps. For simplicity, we assume that all clumps initially have the same mass, $M\sub{cl}$, and the same size, $r\sub{cl}$. Denoting the total clump mass as $M\sub{cl,tot}$, $(M\sub{cl,tot}/M\sub{cl})$ clumps are in the simulation, and each individual clump is modelled with $(M\sub{cl}/m\sub{p})$ particles; i.e., the number of clump particles is given as $M\sub{cl,tot}/m\sub{p}$. The internal structure of giant clumps is uncertain. In this study, we use the NFW density profile, \autoref{eq:nfw_rho}, to describe the internal structure of individual clumps, with the concentration of $c\sub{cl} = r\sub{cl}/r\sub{s} = 1$, so that the density structure of clumps is effectively described by a single power-law of $r^{-1}$, as \nbody particles of individual clumps are sampled in the radial range of $r \leq r\sub{cl}$. While $c\sub{cl}$ is fixed throughout of the paper, the dependence on the density of clumps is studied by varying $r\sub{cl}$. 

The clumps are expected to form through violent disk instability and initialised on the $XY$-plane (i.e., $Z = 0$). We suppose that their initial distribution is described by an exponential disk profile, i.e.,
\begin{equation}
    \frac{dN\sub{cl}}{dR} \propto \exp{(-R/R\sub{d})} R, 
        \label{eq:nclump}
\end{equation}
where $R$ and $R\sub{d}$ are the distance from the centre of the host system to a point on the $XY$-plane and the disk scale length, respectively. Based on \autoref{eq:nclump}, $R$ is drawn by using the acceptance-rejection sampling method with $p(R) \propto dN\sub{cl}/dR$ and a 2D unit vector determines the initial location of the individual clumps. All clumps rotate anticlockwise with a velocity, $[G M\sub{host}(R) / R]^{1/2}$, where $G$ and $M\sub{host}(R)$ are the gravitational constant and the enclosed mass of the host halo within $R$. 

\begin{table}
\begin{center}
\caption{Summary of the simulation parameters. Column 
(1) Simulation ID. 
(2) Total clump mass.  
(3) Mass of individual clumps. 
(4) Clump size. 
(5) Disk scale length. 
\label{tab:params}}
\begin{tabular}{ccccc}
\hline 
(1)         & (2)                     & (3)                 & (4)               & (5)          \\
ID          & $M\sub{cl,tot} [10^9~\msun]$ & $M\sub{cl} [10^9~\msun]$ & $r\sub{cl}$ [kpc] & $R\sub{d}$ [kpc]  \\
\hline
A           & 12                      & 1.5                 &    1              & 10        \\
B           & 12                      & 0.092               &    1              & 10        \\
C           & 12                      & 5.9                 &    1              & 10        \\
D           & 12                      & 1.5                 &    0.25           & 10        \\
E           & 47                      & 1.5                 &    1              & 10        \\
F           & 12                      & 1.5                 &    1              & 5         \\
G           & 12                      & 1.5                 &    1              & 20        \\
H           & 47                      & 5.9                 &    1              & 20        \\
\hline
\end{tabular}
\end{center}
\end{table}
In summary, our giant gas clump model has four parameters, $M\sub{cl,tot}, M\sub{cl}, r\sub{cl}$ and $R\sub{d}$. As the fiducial parameter set, we employ $M\sub{cl,tot} = M\sub{200,host}/256 \approx 1.2 \times 10^{10} \msun$, which corresponds to $\sim 25$ percent of the stellar mass expected for the host halo considered \citep{Moster2018,Behroozi2019}, $M\sub{cl} = M\sub{200,host}/2048 \approx 1.5 \times 10^9 \msun, r\sub{cl} = 1$~kpc, based on the observations of giant clumps \citep[e.g.,][]{Elmegreen2009,ForsterSchreiber2011,Genzel2011}, and $R\sub{d} = 10$~kpc, motivated by the theoretical model by \cite{Mo1998} and \cite{Burkert2016} and the halo spin parameter obtained in cosmological \nbody simulations \citep[e.g.,][]{Bullock2001,Maccio2008,Zjupa2017}. Despite tremendous efforts made by previous studies, there still remains an appreciable amount of uncertainty and scatter in those parameters, and we investigate the dependence on each parameter. \autoref{tab:params} summarises the parameters of the clump simulations.

\subsection{Overview of the dynamical evolution}
\label{ssec:clump_overview}

\begin{figure}
    \begin{center}
        \includegraphics[width=0.45\textwidth]{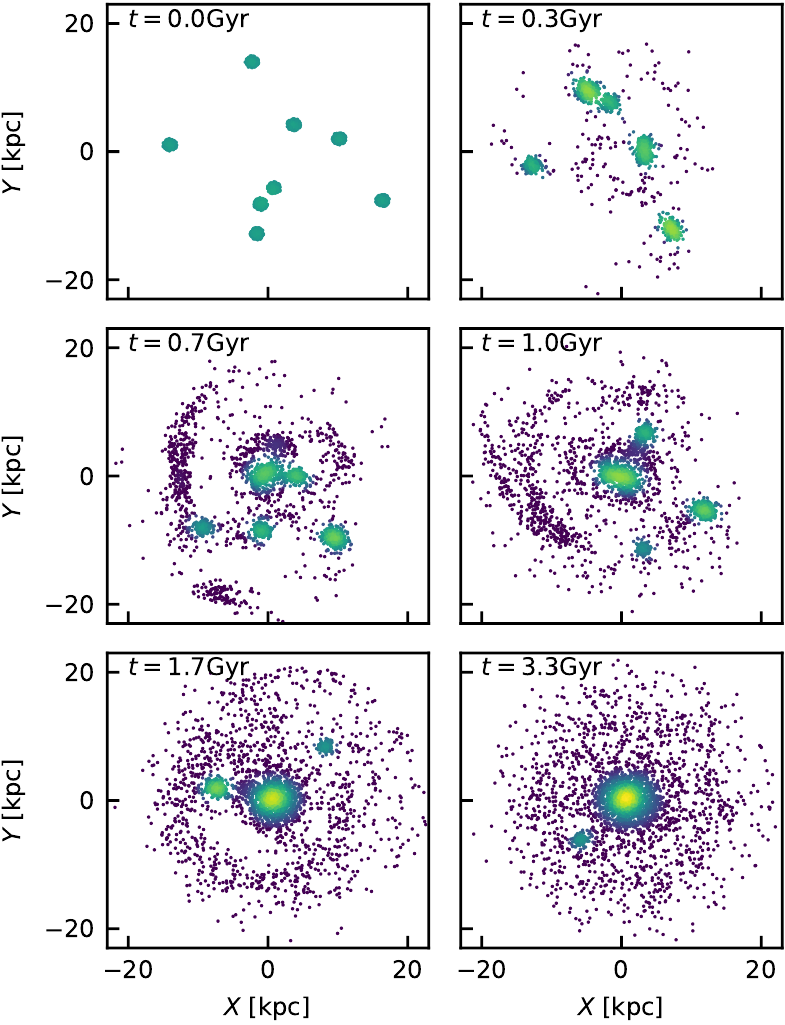}
    \end{center}
        \caption{
        Distribution of clump particles obtained from the run-A. The time of snapshots is shown at the top left corner of each panel. Brighter (Darker) points indicate higher (lower) densities. Clumps are deformed by the tidal force of the host halo and interactions between clumps and sink to the centre of the host halo due to dynamical friction. They form a bulge-like structure at the centre in $\sim 1$~Gyr and it grows with time.  
        \label{fig:clump_image}}
\end{figure}

In this subsection, we study how the internal dynamical structure of the host halo can be altered by orbiting baryonic clumps, ubiquitously formed in gas rich \hiz galaxies. The centre of the system is defined using all particles. \autoref{fig:clump_image} illustrates the global evolution of clumps in the run-A of the fiducial parameter set. The {\sc scipy.stats.gaussian{\_}kde} module \footnote{\url{https://docs.scipy.org/doc/scipy/reference/generated/scipy.stats.gaussian_kde.html}} is used for the density estimation. Clumps initially rotate anti-clockwise on the $XY$-plane (see \autoref{ssec:clump_setup} for the details of the setup) and are deformed by the tidal force of the host halo and interactions between clumps (upper right). The orbit of clumps gradually decays due to the loss of orbital energy and angular momentum driven by dynamical friction. As a result, the majority of clumps sinks towards the centre and forms a bulge-like structure in $\sim 1$~Gyr (middle right). Then, the accumulation of clumps continues and a central bulge-like structure grows with time. The orbital decay timescale of clumps in our simulations corresponds to the upper limit, because the host galaxy would have a higher density at $R < R\sub{d}$ when considering the gas and stellar disks, making the drag force due to dynamical friction stronger. For example, \cite{Ceverino2010} showed a giant clump in a cosmological hydrodynamical simulation merges into a central bulge in $\sim 250$~Myr, consistent with the analytical expectation by \cite{Dekel2009}.

\begin{figure}
    \begin{center}
        \includegraphics[width=0.4\textwidth]{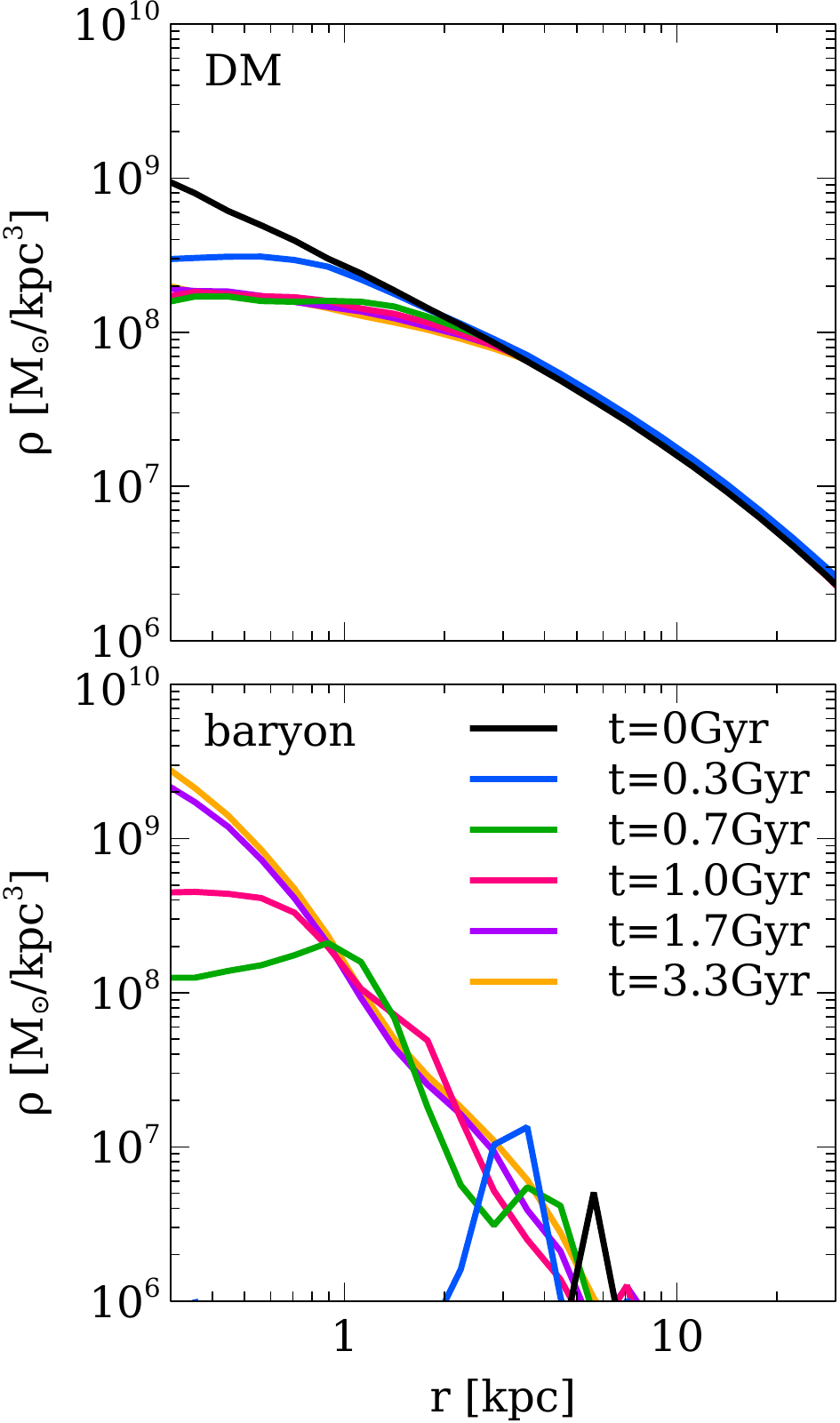}
    \end{center}
        \caption{
        Density profiles obtained from the run-A. In the analysis of the upper and lower panels, particles initially belonging to the host halo and clump particles are included, respectively. The corresponding time of each snapshot is indicated in the legend. The DM central density is reduced and a core is formed at the centre in $\sim 1$~Gyr ({\it upper panel}), as a result of the dynamical heating by giant baryonic clumps. As a back-reaction, clumps sink to the centre and form a bulge-like structure ({\it lower panel}, see also \autoref{fig:clump_image}). 
        \label{fig:clump_run-a}}
\end{figure}

As shown in \autoref{fig:clump_image}, giant baryonic clumps lose their orbital energy and angular momentum and heat up the surround DM particles via dynamical friction. Although the mass of individual clumps is negligible compared to the host mass, the analytical model in \appref{app:ana_model} shows that the dynamical heating could be an efficient heating source in the host centre. We find in the upper panel of \autoref{fig:clump_run-a} that the dynamical heating by baryonic clumps reduces the central DM density and flattens the central cusp out. The density core is formed in a short time scale and grows until $t \sim 1$~Gyr, comparable to the formation timescale of the bulge-like structure. Then, the growth of the density core is halted and the cored structure persists until the end of the simulation ($t \sim 3$~Gyr). The saturation of the core size is understood as the consequence of the depletion of the heating source. The majority of clumps has lost their orbital energy and angular momentum by $t \sim 1$~Gyr. While additional baryons are accreted in the subsequent phase, the amount is small (lower panel). We emphasise that only baryons initially belonging to giant clumps are regarded as baryons in the simulation, while a large fraction of baryons is, in fact, distributed in other components, such as the bulge and stellar and gas disks. Since the formation and accretion of new clumps during simulations are not considered, orbiting clumps working as dynamical heating sources are depleted and the growth of the DM density core is halted. Originated by the stochastic process to set the initial configuration of the clump simulations (see \autoref{ssec:clump_setup}), they have non-zero realisation-to-realisation variance. We confirm that the resultant density profile hardly depends on the stochastically originated variances. Note that in reality the orbital energy and angular momentum of baryonic clumps are re-distributed not only to the DM but also to the galactic disk of the host. Therefore, the impact of the dynamical heating obtained in clump simulations should be taken as the upper limit.

\subsection{Dependence on clump properties}
\label{ssec:clump_dependence}

\subsubsection{Density profile}
\label{sssec:clump_rho}

\begin{figure*}
    \begin{center}
        \includegraphics[width=0.95\textwidth]{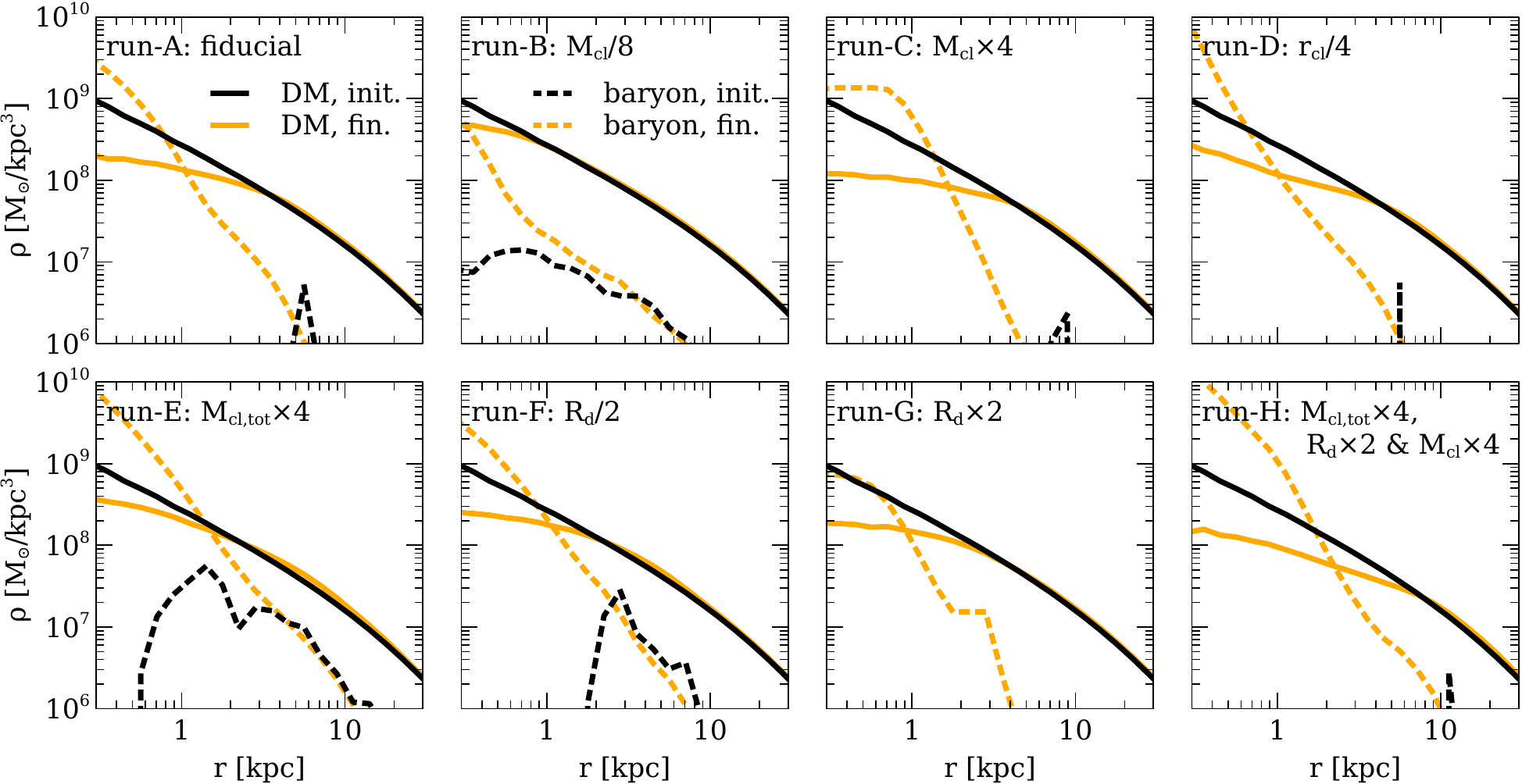}
    \end{center}
        \caption{
        Density profiles obtained from clump simulations. Each panel shows results from a clump simulation and the ID and feature of each run are indicated (see also \autoref{tab:params} for adopted parameter sets). Black and orange lines show the initial and final snapshots of the simulations. Solid and dashed line styles represent the density profile of the DM and baryonic components, respectively. Simulations are consistent with each other, qualitatively, i.e., dynamical friction drives the condensation of baryons in the centre and the re-distribution of energy and angular momentum flattens the DM central cusp out. See the main text for the detailed discussion. 
        \label{fig:clump_rho}}
\end{figure*}

The size of the resultant DM core ($\sim 2$~kpc) in the run-A is, in fact, smaller than the observationally suggested ($\sim 10$~kpc). In \autoref{fig:clump_rho}, we study how the resultant density profile depends on the parameter choice. Black- and orange lines show the initial ($t = 0$~Gyr) and final state ($t = 3.3$~Gyr) in the simulations, while solid and dashed line styles represent the DM and baryonic components, respectively. The run ID and the feature of the parameter choice (see also \autoref{tab:params}) are indicated at the top left corner of each panel. 

We first study the dependence on the mass of individual clumps, $M\sub{cl}$, with the runs-B and -C, in which a smaller and larger $M\sub{cl}$ is adopted respectively, while other parameters are unchanged. Comparing to the run-A, $M\sub{cl}$ in the run-B is smaller by a factor of eight. Since the total clump mass is the same, the number of clumps is eight times larger in the run-B, which is reflected to the initial smoother distribution of the baryon (dashed black). We find that in the run-B the reduction of the central DM density is limited and the core size is smaller than in the run-A (solid orange). Because dynamical friction works inefficiently for low mass clumps, as indicated by Chandrasekhar's dynamical friction formula \citep{Chandrasekhar1943} that the drag force is proportional to the mass squared of the object decelerated by dynamical friction (baryonic clumps), the efficiency of the dynamical heating gets lower. As a consequence, the condensation of the baryon at the centre is suppressed and the baryon distribution remains more extended than in the run-A. In the run-C, the resultant DM density is comparable to that in the run-A, while $M\sub{cl}$ is four times larger. A dominant fraction of the baryon mass condensates in the centre after dynamically heating DM particles, as observed in the run-A. Because of the depletion of the subsequently sinking clumps, the growth of the DM core is halted. While only two clumps are included in the run-C, we confirm that the realisation-to-realisation scatter in the resultant density profile is negligible.

The run-D contains more compact clumps of $r\sub{cl}=0.25$~kpc (cf. $r\sub{cl}=1$~kpc in other runs), while other parameters are the same as in the run-A. The initial position and velocity of clumps in the run-D are identical to those in the run-A. We find that in the run-D the core size of the DM density distribution gets larger and the baryon density gets higher in the centre. Denser clumps are more resilient to the tidal force of the host halo and experience less tidal mass loss. This keeps the efficiency of dynamical friction high for a longer time. As a consequence, the DM central cusp is heated up more significantly and a larger amount of baryons sinks to the centre. 

We increase the total clump mass, $M\sub{cl,tot}$, by a factor of four in the run-E and find that the core formation scenario of inspiralling clumps induces competing phenomena, one of which induces heating whereas the other induces cooling. $M\sub{cl,tot}$ in the run-E is comparable to the stellar mass expected by the theoretical models of galaxy formation and evolution \citep[e.g.,][]{Moster2018,Behroozi2019}. In the run-E, the high central baryon density is obtained since the total baryon mass is large and clumps are massive enough to keep a high efficiency of dynamical friction. While one may expect a larger core size since the DM cusp should be heated up strongly by the clumps, it is actually smaller than in the run-A, due to the counter effect caused by the condensed baryons in the centre \citep{El-Zant2004,Inoue2011}. The bulge-like baryonic structure pulls back the DM to the centre and effectively cools the central part of the system down \citep{Blumenthal1986,Gnedin2004,Zemp2012}. The competition of the dynamical heating and cooling effects of baryons determines the resultant DM density structure. The result from the run-E indicates that a larger total clump mass enhances both the heating and cooling effects. When the mass of the bulge-like structure is large, the cooling effect works strongly and suppresses the reduction of the DM central density. 

Next, we study the dependence on the size of the initial clump distribution controlled by a parameter, $R\sub{d}$, with the runs-F and -G. The initial clump distribution is more concentrated with a halved $R\sub{d}$ in the run-F, while the other parameters are the same as those in the run-A. Because clumps are initialised at smaller galactocentric radii, they have smaller energy and angular momentum compared with those of clumps in the run-A. Thus the dynamical heating is weaker and the resultant core size is smaller than in the run-A. We set a more extended clump distribution with a doubled $R\sub{d}$ and a large amount of orbital energy and angular momentum of clumps is available to heat the central DM cusp up in the run-G. However, the resultant DM core size is smaller than that found in the run-A (solid orange), because the baryon mass sank to the centre, i.e., clumps released their orbital energy and angular momentum, is smaller as indicated by the lower central baryon density (dashed orange). Due to the large orbital energy and angular momentum, their orbits are large and each clump heats the halo centre up less efficiently by the end of the simulation. 

These results imply that there would be a sweet spot in the four dimensional parameter space ($M\sub{cl,tot}$, $M\sub{cl}$, $r\sub{cl}$, and $R\sub{d}$) to create a large DM core. As demonstrated, sinking baryonic clumps heat the DM cusp up and can flatten it out. However, they can also cause a negative dynamical cooling effect after sinking. The condensed baryons pull DM back to the centre and suppress the growth of the density core. The resultant DM density profile depends on the competition of the effects of the dynamical heating and cooling. After some parameter survey, we find a parameter set to form a DM core with a size of $\sim 10$~kpc, as suggested by observations (run-H). Forming a larger DM core with sinking baryonic clumps may be possible, while the required parameters may be unfavoured by the standard galaxy formation models. 

\begin{figure}
    \begin{center}
        \includegraphics[width=0.45\textwidth]{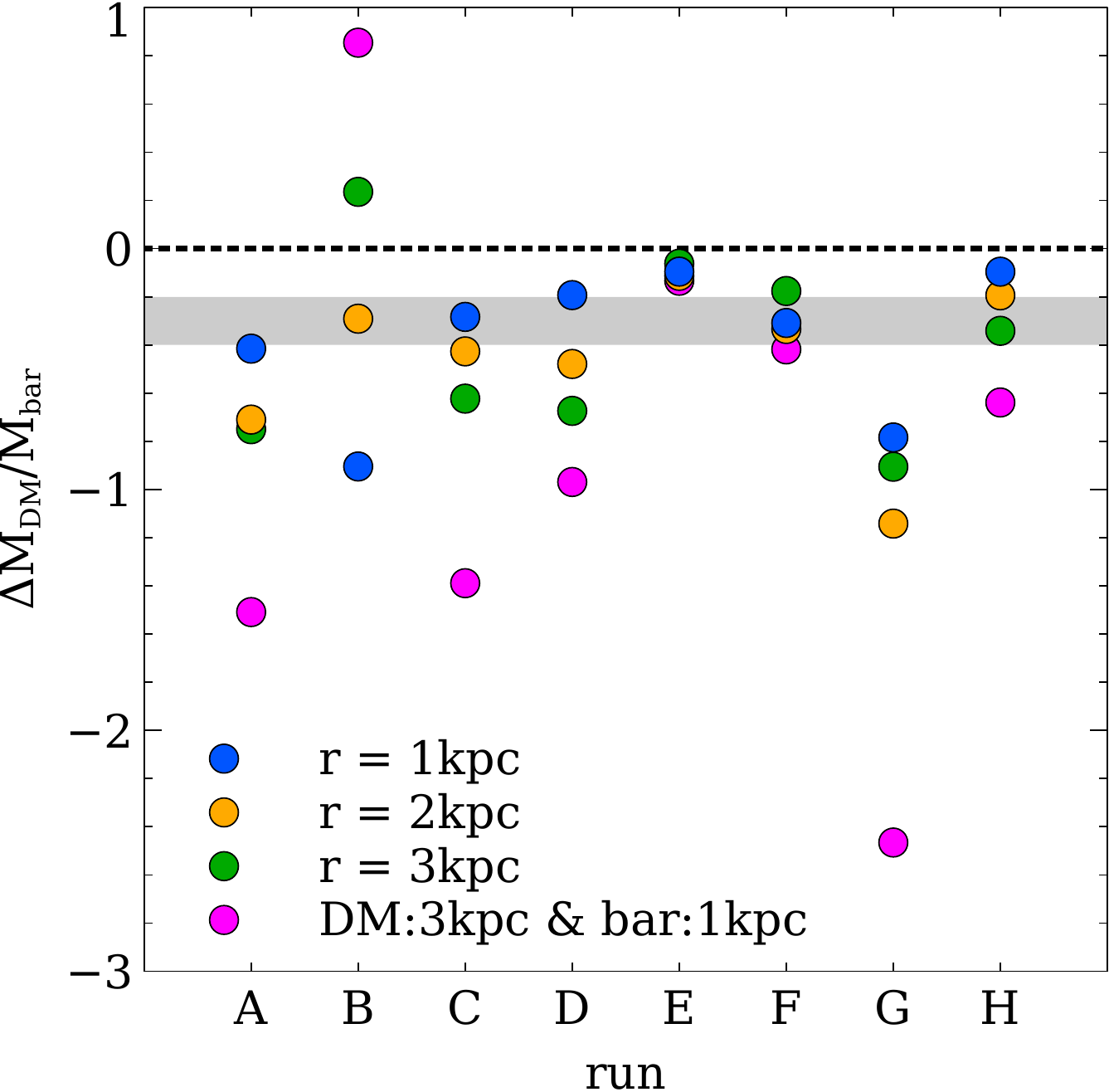}
    \end{center}
        \caption{
         DM mass removal efficiency, $\Delta M\sub{DM}(r)/M\sub{bar}(r)$, measured at $r=1$ (blue), 2 (orange) and 3\,kpc (green), in each clump simulation. 
         In the analysis of magenta circles, $\Delta M\sub{DM}$ is measured at $r=3$~kpc, while $M\sub{bar}$ is measured at $r=1$~kpc.  
         Shaded band indicates the ratio of the difference between the observationally inferred and theoretically expected DM masses to the bulge mass of \hiz star forming galaxies, obtained by \citet{Genzel2020}. Horizontal dotted black line marks the DM mass removal efficiency of zero for reference. Most of points are below the horizontal dotted black line, indicating that the dynamical heating of giant clumps reduces the central DM mass, while the efficiency depends on the clump properties and the radius to measure it.
        \label{fig:heat_eff}}
\end{figure}

\cite{Genzel2020} found that the DM mass within the optical light effective radius of \hiz massive star forming galaxies is smaller than the $\Lambda$CDM prediction by $\sim 30$ percent of the bulge mass. As demonstrated above, a bulge-like structure can be formed as giant clumps sink to the galactic centre due to dynamical friction. The DM mass removed from the centre would correlate with the accreted baryon mass, if the DM core is originated by the dynamical heating of giant clumps. We compare the change in DM mass, $\Delta M\sub{DM}(r) \equiv M\sub{DM,fin}(r) - M\sub{DM,ini}(r)$, where $M\sub{DM,ini}(r)$ and $M\sub{DM,fin}(r)$ are the DM mass profile at the beginning ($t=0$) and end ($t=3.3$~Gyr) of clump simulations, to the baryon mass at the end of the simulations, $M\sub{bar}(r)$. As shown in \autoref{fig:clump_rho}, the baryon mass initially enclosed within the central a few kpc is negligible compared with the baryon mass enclosed within the same radial range at the final state of the simulations. Thus, $M\sub{bar}(r)$ effectively corresponds to the accreted baryon mass, and the ratio, $\Delta M\sub{DM}(r)/M\sub{bar}(r)$, indicates the efficiency of the DM mass removal by the dynamical heating of accreted baryons. 

In \autoref{fig:heat_eff}, we compare the DM removal efficiency in each clump simulation to the observationally inferred ratio of the DM mass deficit to the bulge mass (grey shaded band). While \cite{Genzel2020} inferred the DM mass deficit at the optical light effective radius of the \hiz galaxies, the radius to measure the DM removal efficiency is parametrised in our analysis. Since our simulation model lacks some baryonic components, such as the gas and stellar disks, it is difficult to adequately define the effective radius of the galaxy model. Combining the stellar mass to halo mass relation by \cite{Behroozi2019} and the size-mass relation by \cite{vanderWel2014}, we expect the effective radius of the stellar body of the galaxy to be about 3~kpc for the host model with $M\sub{200,host} = 3 \times 10^{12}~\msun$, while the bulge is typically more compact than galactic disks by a factor of a few \citep{Graham2001}. In the analyses of blue, orange and green circles, both $\Delta M\sub{DM}$ and $M\sub{bar}$ are measured at $r=1$, 2 and 3~kpc. We also perform another analysis in which $\Delta M\sub{DM}$ is measured at $r=3$~kpc, while $M\sub{bar}$ is measured at $r=1$~kpc (magenta).

\autoref{fig:heat_eff} clearly shows that $\Delta M\sub{DM}(r)/M\sub{bar}(r)$ is negative, except when measuring $\Delta M\sub{DM}$ at $r=3$~kpc in the run-B, indicating that the dynamical heating of giant clumps effectively reduces DM mass from the galactic centre. We find that the DM removal efficiency varies from ten to 250 percent, depending on the clump properties and the radius to measure the efficiency. When measuring both $\Delta M\sub{DM}$ and $M\sub{bar}$ at the same radius (blue, orange or green), the runs-B, -C, -F and -H have the DM mass deficit to baryon mass ratio consistent with the observations, implying that the observed DM mass deficit could be originated by the dynamical heating of giant baryonic clumps. In the analysis of magenta circles, the absolute value of the DM removal efficiency is amplified compared to green circles as $M\sub{bar}$ is small, and only the run-F is consistent with the observations. As our idealised simulation model does not allow comparison of the simulation results to observations in a self-consistent way, analysing cosmological hydrodynamical simulations is a potential avenue for further exploring the role of giant clumps in the formation and evolution of \hiz galaxies.

\subsubsection{DM mass fraction}
\label{sssec:clump_fdm}

\begin{figure*}
    \begin{center}
        \includegraphics[width=0.95\textwidth]{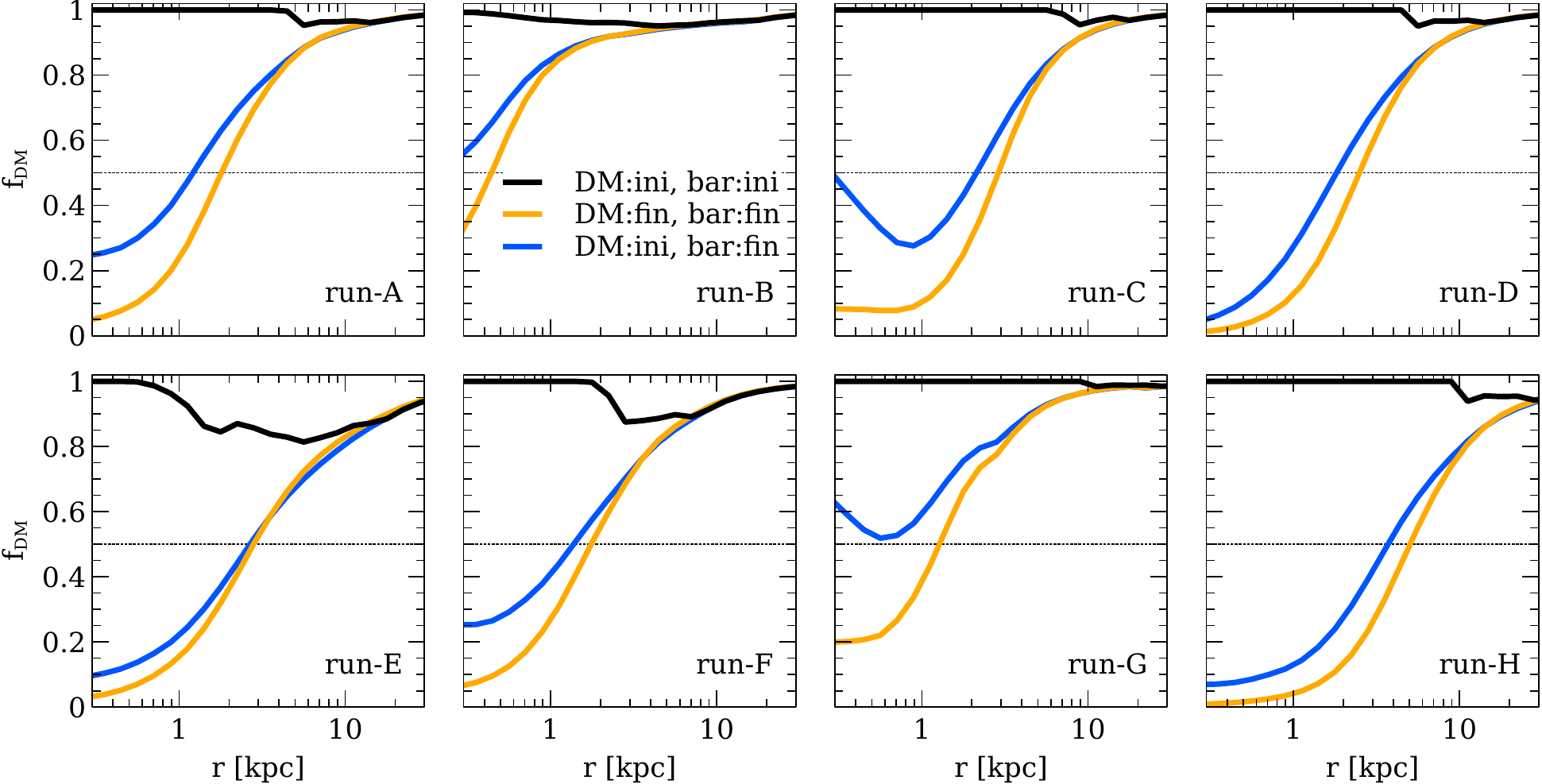}
    \end{center}
        \caption{
        DM mass fraction as a function of $r$, $f\sub{DM}(r)$. In each panel, three combinations of the initial and final enclosed mass profiles are shown as indicated: black and orange lines depict the initial and final states, respectively. To disentangle the impact of the DM core formation from that of the baryon condensation, blue line adopts the initial DM profile having a central cusp. Horizontal black dotted line marks $f\sub{DM}=0.5$ for reference. The baryon condensation reduces the central DM mass fraction (blue). The DM core formation further boosts the reduction of $f\sub{DM}$ (orange).
        \label{fig:clump_fdm}}
\end{figure*}

In \autoref{fig:clump_fdm}, we study the DM mass fraction defined as
\begin{equation}
    f\sub{DM}(r) \equiv \frac{M\sub{DM}(r)}{M\sub{DM}(r)+M\sub{bar}(r)},
        \label{eq:fdm}
\end{equation}
where $M\sub{DM}(r)$ and $M\sub{bar}(r)$ are the DM and baryon mass enclosed within $r$, respectively. As depicted in \autoref{fig:clump_rho}, clumps can reduce the DM mass fraction at the galaxy centre i) by reducing the DM mass (DM core formation); and i\hspace{-.1em}i) by increasing the baryon mass (baryon condensation). To disentangle the two mechanisms, we show $f\sub{DM}(r)$ based on the initial DM distribution with the central cusp (blue) and the final cored distribution (orange), while the final baryon profile is employed for the both. For comparison, the black line shows the initial $f\sub{DM}(r)$, employing the initial DM and baryon distributions. 

\autoref{fig:clump_fdm} shows that in clump simulations the combination of the baryon condensation and the DM core formation significantly reduces $f\sub{DM}$ at $r < 1$~kpc (orange), except for the run-B in which either mechanism does not work well due to the low efficiency of dynamical friction. Comparing the orange line to the blue line, we find that the baryon condensation is the primary mechanism to reduce the central $f\sub{DM}$. In the runs-A, -C, -D, -G and -H in which a large DM core ($\ga 2$~kpc) has been formed by the dynamical heating of clumps, the core formation boosts the reduction of the central $f\sub{DM}$ substantially and expands the radial range of the baryon domination (at which the blue or orange line is below the horizontal dotted black line) by a factor of $\sim$ two. 

Our simulations demonstrate that giant baryonic clumps can effectively reduce $f\sub{DM}$ at the central kpc of the \hiz massive galaxy. However, it is still hard to conclude that they fully explain the low $f\sub{DM}$ ($\sim 0.2$) inferred at the effective radius ($\ga 3$~kpc) of massive galaxies at $z \sim 2$ \citep{Genzel2020}. Since the simulations do not include other baryon components, such as a pre-existing bulge and stellar and gas disks, $f\sub{DM}$ shown in \autoref{fig:clump_fdm} corresponds to the upper limit. The existence of those components can reduce $f\sub{DM}$ by increasing $M\sub{b}$, while it could suppress the core formation due to the negative dynamical cooling effect. Also, the inclusion of supernova feedback and hydrodynamics could alter the evolution of individual clumps. More detailed studies based on the full hydrodynamical cosmological simulations would be needed to test this scenario.

\section{Summary}
\label{sec:summary}

Recent observations found peculiar rotation curves of star forming massive galaxies at $z \sim 2$. The rotation velocity of the galactic gas peaks around the effective radius of the galaxies and declines outward. \cite{Genzel2020} inferred that DM accounts for only 10-20 percent of the dynamical mass within the effective radius of those galaxies and thus the galaxies are dynamically dominated by baryons. They also interpreted the declining rotation curves as the indication of a central DM core rather than the cuspy structure predicted by cosmological \nbody simulations based on the concordance $\Lambda$CDM paradigm. While the recursive potential fluctuation driven by supernova feedback has a potential to flatten the cusp in dwarf galaxies, it would not be powerful enough to make a core in \hiz massive galaxies. Therefore, other mechanisms heating the cusp are needed to understand the formation of the baryon dominated galaxies having a DM density core. 

\cite{Dekel2021} considered a combination of a major galaxy merger and AGN feedback as a heating source to reduce the DM mass in the centre of \hiz massive galaxies. The satellite galaxy is supposed to be significantly contracted due to the compaction effect prior to the merger event. The dynamical heating by the satellite makes DM particles initially belonging to the host vulnerable to the subsequent dynamical heating induced by AGN feedback and helps make a large DM core. However, our merger simulation demonstrates that a large amount of DM mass is supplied by the dense satellite. Since the compaction effect requires the satellite to be massive enough ($M\sub{200,sub} \ga 10^{11.3}~\msun$), dynamical friction decays the satellite orbit in a few Gyr and the satellite replenishes the central DM density, even if a DM core is formed temporarily.

This result motivates us to consider another heating source, giant baryonic clumps, ubiquitously observed in \hiz galaxies and expected to be formed through the fragmentation of the galactic disk. The clumps are dense enough to survive in the strong tidal force field at the host centre and thus have a potential to flatten out the central cusp of the host. We study how giant baryonic clumps can alter the density structure of the host by using clump simulations varying the clump properties, such as their mass and size, and radial distribution. Our clump simulations find that the orbit of baryonic clumps decays due to dynamical friction and as a consequence they form a bulge-like structure in the galaxy centre within a few Gyr. The re-distribution of orbital energy and angular momentum from clumps to the host halo creates the DM density core at the host centre. These two mechanisms both work to decrease the central DM dominance and become efficient when clumps are massive enough, $\ga 10^9~\msun$, due to the high efficiency of dynamical friction. Our clump simulations demonstrate that the baryon condensation works primarily to reduce the DM mass fraction in the central kpc scale. When a DM core is formed by clumps, the baryon dominated range is expanded to 2-5~kpc, comparable to the effective radius of \hiz massive galaxies. We also find that the DM removal efficiency in clump simulations is consistent with the corresponding observationally inferred quantity, assuming that the bulge is formed through the accumulation of giant clumps. Based on these results, we conclude that giant baryonic clumps offer a promising explanation for the origin of the baryon domination and the large DM core in massive \hiz galaxies.

We note that our clump simulations neglect some relevant processes, such as feedback from stars and AGN and gas supply from large scale structures, possibly altering the clump properties and evolution. First of all, feedback could suppress the formation of massive clumps having a mass of $\sim 10^9 ~ \msun$, a key to form a large DM core. The demographics and distribution of clumps depend on sub-grid physics models of simulations \citep{Tamburello2015,Mayer2016,Mandelker2017}. Second, clumps would be sequentially formed in the galactic disk, while our simulations include only pre-set clumps. In this sense, supply of gas from the large scale structure may play an important role in determining the global properties of clumps within the galaxy \citep{Dekel2009_stream}. In addition, some fraction of clumps can also be supplied by galaxy mergers \citep{Mandelker2014}. 

Our idealised clump simulations demonstrate that the efficiency of the dynamical heating by giant clumps and the resultant baryon domination strongly depend on clump properties, demographics and distribution. Future work should focus on investigating the impact of dynamical heating by giant clumps in \hiz galaxies in high-resolution cosmological hydrodynamical simulations. To resolve the formation of giant clumps and its dynamical heating effects, the simulations must achieve the spatial resolution of $\la 10$~pc  \citep[e.g.,][]{Ceverino2010,Behrendt2019}. For example, the {\tt VERA} cosmological zoom-in hydrodynamical simulations \citep{Ceverino2014,Zolotov2015,Mandelker2017}, which reproduce the observed clumpiness of galaxies and the star formation rate of clumps, could make an interesting sample to investigate the role of giant clumps in the formation and evolution of \hiz galaxies.

\section*{Acknowledgements}

The authors thank Andreas Burkert and Avishai Dekel for useful discussions and the anonymous referee for providing insightful comments. GO acknowledges the Waterloo Centre for Astrophysics Fellowship for the support. DN acknowledges support from Yale University. Numerical simulations were performed on the Graham cluster operated by Compute Canada (\url{www.computecanada.ca}).

\section*{Data Availability}
The data and code underlying this article will be shared on reasonable request to the corresponding author.



\bibliographystyle{mnras}
\bibliography{highz_core}


\appendix

\input{./ana_model.tex}


\bsp	
\label{lastpage}
\end{document}

%% file: ana_model.tex
\section{Analytical model of dynamical heating}
\label{app:ana_model}

Let us suppose that a substructure (e.g., a satellite galaxy or a giant baryonic clump) is orbiting in a larger spherical host following the NFW density profile (\autoref{eq:nfw_rho}). Several physical mechanisms transfer the orbital energy and angular momentum from the substructure to the host system and alter the dynamical structure of the host. Here, we consider two mechanisms: dynamical friction (\autoref{ssec:dynamical_friction}) and tidal shock (\autoref{ssec:tidal_shock}). Note that the contribution by self-friction, i.e., another type of drag force provided by the mass stripped from the substructure  \citep{Fujii2006,Fellhauer2007,vandenBosch2018b,Ogiya2019}, is subdominant compared to dynamical friction \citep{Miller2020}, and thus we neglect it.

\subsection{Dynamical friction}
\label{ssec:dynamical_friction}

\begin{figure*}
    \begin{center}
        \includegraphics[width=0.9\textwidth]{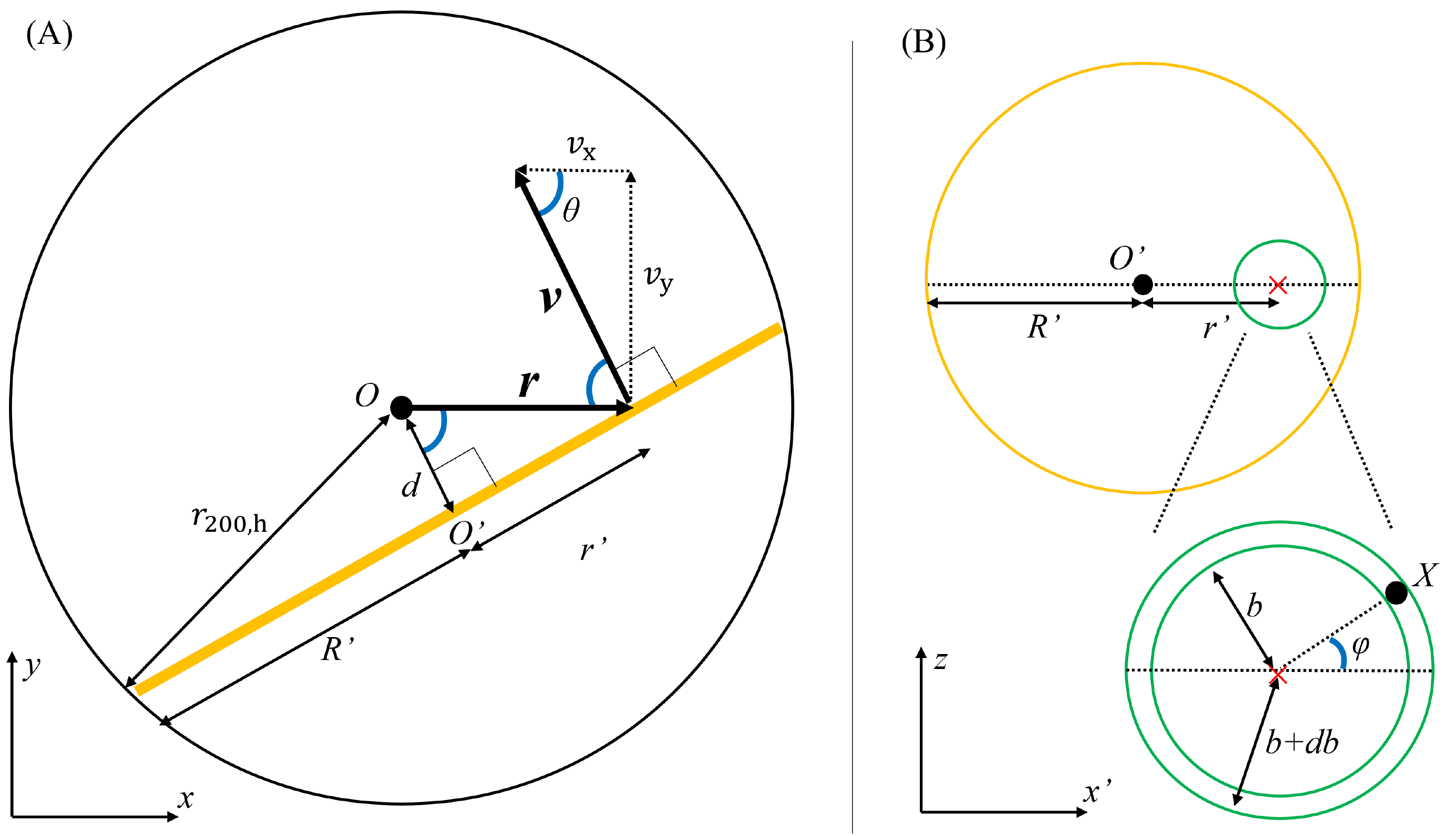}
    \end{center}
        \caption{
        Schematic picture of the kernel model. 
        A plane parallel to $z$-axis is shown as the orange line and orange circle in Panels-A and -B, respectively, in order to evaluate the efficiency of the energy transfer from a substructure to the host via DF. 
        {\it Panel-A:} The substructure is located on the plane indicated by the orange line, which is perpendicular to the orbital plane of the substructure orbiting within a host system ($xy$-plane) and the velocity vector of the substructure (black arrow, ${\bf v}$). 
        {\it Panel-B:} A ring of the impact parameter, $b$, (green) on the plane introduced in Panel-A (orange) is considered. The red cross indicates the location of the substructure on the plane. 
        See the main text for the details.
        \label{fig:geometry}}
\end{figure*}

The gravity of a substructure orbiting within a larger host system induces enhancements in the density field of the host (density wakes) behind the substructure \citep[e.g.,][]{Weinberg1989,Ogiya2016,Vasiliev2022}. The wakes pull back the substructure, causing a drag force on it. This process is known as dynamical friction (DF; \citealt{Chandrasekhar1943}) and drives the orbital decay of substructures \citep[e.g.,][]{Lacey1993,vandenBosch1999,Taylor2001,Jiang2008}. We adopt the host-centric coordinates and suppose the substructure as a point mass with a mass of $M$. According to \cite{Chandrasekhar1943}, the drag force of DF is 
\begin{equation}
    {\bf F}\sub{DF}({\bf r}) = -4 \pi G^2 M^2 \ln{\Lambda} \frac{\rho({\bf r})f(v) }{v^3({\bf r})} {\bf v}({\bf r}),
        \label{eq:adf}
\end{equation}
where ${\bf r}$ and ${\bf v}$ are the position and velocity vectors of the substructure. The Coulomb logarithm and the mass density of the host are represented as $\ln{\Lambda}$ and $\rho$, respectively. Only particles belonging to the host with velocities less than $v$ are expected to contribute to DF, and $f(v)$ is the fraction of the mass of particles participating in the DF process. Assuming the Maxwell-Boltzmann velocity distribution with a velocity dispersion of $\sigma$, $f(v)$ is given by
\begin{equation}
    f(v) = {\rm erf}\biggl ( \frac{v}{\sqrt{2}\sigma} \biggr )  - \sqrt{\frac{2}{\pi}}\frac{v}{\sigma}\exp{\biggl (- \frac{v^2}{2\sigma^2} \biggr )}, 
        \label{eq:fv}
\end{equation}
where $\sigma(r) \equiv [GM\sub{h}(r)/r]^{1/2}$ and $M\sub{h}(r)$ is the enclosed mass profile of the host. When assuming a spherical host system, the velocity of a substructure with a specific orbital energy of $\epsilon$ is a function of $r$,
\begin{equation}
    v(r) = \sqrt{2[\epsilon- \Phi(r)]},
        \label{eq:v}
\end{equation}
where $\Phi(r)$ is the gravitational potential profile of the host.

\subsubsection{DF as a local process}
\label{sssec:df_local}

We evaluate the orbital energy of the substructure transferred to host particles. The work done by the DF drag force locally would correspond to the amount of the locally deposited energy. The displacement of the substructure in a time interval of $dt$ is $ds = v dt$. This yields the work done by the DF drag force in a spherical shell of $[r : r + dr]$, 
\begin{equation}
    dE\sub{DF}(r) = \biggl | F\sub{DF}(r) \frac{v(r)}{v\sub{r}(r)} \biggr | dr,
        \label{eq:dE_df}
\end{equation}
where $dr$ is the radial displacement in $dt$, and we used $dt(r) = dr / |v\sub{r}(r)|$. The radial velocity of the substructure with a specific energy of $\epsilon$ and a specific angular momentum of $l$ at $r$ is
\begin{equation}
    v\sub{r}(r) = \sqrt{2[\epsilon- \Phi(r)] - l^2/r^2}.
        \label{eq:vr}
\end{equation}
The specific energy deposited in the mass shell is defined as $d \epsilon\sub{DF}(r) \equiv dE\sub{DF}(r)/[4 \pi \rho(r)r^2 dr]$. To study the radial range of the host where the dynamical structure can be altered by DF, we compare $d \epsilon\sub{DF}(r)$ and the gravitational potential, $\Phi(r)$:
\begin{equation}
    \frac{d\epsilon\sub{DF}(r)}{|\Phi(r)|} = \frac{G^2 M^2 f(v) \ln{\Lambda}}{2 r^2 |\Phi(r)|} [\epsilon-\Phi(r)]^{-1/2} \biggl [\epsilon-\Phi(r)-\frac{l^2}{2r^2} \biggr ]^{-1/2}.
        \label{eq:depsdf_phi}
\end{equation}
Note that DF also reduces the angular momentum of the substructure, where the specific angular momentum lost from the substructure in $dt$ is given by $dl =  dt |{\bf r} \times {\bf F}\sub{df}|/M$. We denote the angle between ${\bf r}$ and ${\bf v}$ as $\Theta$. Using $dt(r) = dr / |v\sub{r}(r)|$ and $\sin{\Theta} = v\sub{t}/v$, where $v\sub{t}=l/r$ is the tangential velocity of the substructure, the specific angular momentum loss at $[r:r+dr]$ is given by
\begin{equation}
    dl(r) = \biggl | \frac{F\sub{DF}(r)}{M} \frac{v\sub{t}(r)}{v\sub{r}(r)} \biggr | \frac{r}{v(r)} dr .
        \label{eq:dL}
\end{equation}

\subsubsection{DF as a global process}
\label{sssec:df_global}

DF arises as a global phenomenon, since gravity is a long range force. Thus the orbital energy of the substructure will be deposited to the host system with some spread. The drag force of DF is a cumulative impact of hyperbolic encounters (i.e., gravitational interactions of unbound orbits) between the substructure and a host particle having a mass of $m$. The relative velocity before the interaction is denoted as ${\bf v}\sub{0} = {\bf v}\sub{M} - {\bf v}\sub{m}$, where ${\bf v}\sub{M}$ and ${\bf v}\sub{m}$ are the velocity vectors of the substructure and the host particle, respectively. We denote the component of the relative position vector of the substructure and host particle, perpendicular to ${\bf v}\sub{0}$, the so-called impact parameter, as $b$. The velocity perturbation caused by the interaction is given as $\Delta {\bf v} \equiv \Delta {\bf v}\sub{M} - \Delta {\bf v}\sub{m} = (\Delta v\sub{\|}, \Delta v\sub{\perp})$, where $\Delta v\sub{\|}$ and $\Delta v\sub{\perp}$ are the components parallel and perpendicular to ${\bf v\sub{0}}$, respectively. After some algebra, we derive 
\begin{equation}
    \Delta v\sub{\|} = \frac{2 v\sub{0}}{1 + (b/b\sub{90})^2}
        \label{eq:dvpara} 
\end{equation}  
\begin{equation}
    \Delta v\sub{\perp} = \frac{2 v\sub{0} (b/b\sub{90})}{1 + (b/b\sub{90})^2}.
            \label{eq:dvperp}
\end{equation}
Here, $v\sub{0} = |{\bf v}\sub{0}|$ and $b\sub{90} \equiv G(M+m)/v\sub{0}^2$ is the impact parameter causing a deflection of 90 degrees. The detailed derivation of the analytical solution of the hyperbolic encounter is found in \cite{Binney2008}. 

The substructure will interact with host particles with a broad range of values in $b$ and ${\bf v}\sub{m}$, when moving through the host system. Momentum conservation tells us that the velocity kick that the substructure receives is $\Delta {\bf v}\sub{M} = m \Delta {\bf v}/(M+m)$ for a single interaction. The drag force by DF in \autoref{eq:adf} is derived by integrating the parallel velocity perturbation (\autoref{eq:dvpara} multiplied by $m/(M+m)$) over $b$ and ${\bf v}\sub{m}$ \citep[e.g.,][]{Mo2010}. On the other hand, the net velocity change of the perpendicular velocity component is assumed to be small (it is zero in the limit where host particles are homogeneously distributed with an infinite extent). Consequently, energy is transferred from the substructure to host particles through the change in $v\sub{m,\|}$. The change in the kinetic energy of a host particle is therefore given by 
\begin{equation}
     dK \propto v\sub{m,\|}^2 - v\sub{0,m,\|}^2 = 2v\sub{0,m,\|}\Delta v\sub{m,\|} + \Delta v\sub{m,\|}^2,
\label{eq:dK}
\end{equation}
where $v\sub{m,\|} = v\sub{0,m,\|} + \Delta v\sub{m,\|}$ and $v\sub{0,m,\|}$ is the parallel component of the velocity vector of the host particle before the interaction with the substructure. By integrating $dK$ over ${\bf v}\sub{m}$, the contribution of the first term is zero when the velocity dispersion of host particles is isotropic.  Therefore, we consider only the second term in what follows.

Next, we consider the efficiency of DF heating in the density-stratified hosts. A shortcoming of the derivation of Chandrasekhar's formulation of DF is the assumption that host particles are homogeneously distributed with an infinite extent. Clearly, it is not the case for realistic astrophysical systems, such as galaxies and DM haloes. The number density of host particles (e.g., stars and DM) depends on the location in the host system and the extent of the host particle distribution is finite. Closing the gap is essential in evaluating how much of energy is re-distributed to a given radial shell of the host system. Specifically, we count the number of host particles interacting with the substructure with an impact parameter, $b$, on a plane on which the substructure is located and perpendicular to the velocity vector of the substructure. The integration of $dK$ over $b$ corresponds to the dynamical heating of DF.

\autoref{fig:geometry} describes the geometry of the system. A substructure orbits within a spherical host system with a virial radius, $r\sub{200,h}$. In the panel-A, we define the orbital plane of the substructure be the $xy$-plane. The position and velocity vectors of the substructure in the host-centric coordinate are defined as ${\bf r} \equiv (r, 0, 0)$ and ${\bf v} \equiv (v\sub{x}, v\sub{y}, 0)$. Given $\epsilon$, $l$ and $r$ as well as the structural parameters of the host, $\rho\sub{s}$ and $r\sub{s}$, we can compute $|v\sub{y}|=l/r$ and $|v\sub{x}|$ using \autoref{eq:v}. Here, we consider a plane perpendicular to the $xy$-plane and ${\bf v}$ and the substructure is on it (orange). The angle between the $x$-axis and ${\bf v}$ is denoted as $\theta$ and is equal to the angle between ${\bf r}$ and ${\bf v}$. The angle between ${\bf r}$ and a vector pointing from the origin (centre of the host) to the plane is also $\theta$ (blue). The distance from the origin to the plane is given as $d = r \cos \theta = r |v\sub{x}|/v$, where $v = |{\bf v}|$. The plane is a circle with a radius of $R' = \sqrt{r\sub{200,h}^2 - d^2}$, and the distance from the centre of the circle $O'$ to the substructure is given by $r' = r \sin \theta = r |v\sub{y}|/v$.

The panel-B of \autoref{fig:geometry} shows a ring with a radius of $b$ (green) around the substructure (red cross) on the plane introduced above (represented as the orange line in the panel-A). Note that the coordinate on the plane is defined by the $x'$- and $z$-axes. The former is derived by rotating the $x$-axis by $(\pi/2 - \theta)$~radian about the $z$-axis. The distance from the centre of the host, $O$, to a point on the ring, $X$ is given by
\begin{equation}
    D = \sqrt{d^2 + r'^2 + b^2 + 2r'b \cos \varphi}, 
        \label{eq:D}
\end{equation} 
where $\varphi$ is the angle between the $x'$-axis and the vector pointing $X$ from $O'$. Since the area element in the ring is $dA = b db d\varphi$ and the substructure passes through the ring, it interacts with $dN \propto \rho(D) dA \propto \rho(D)b$ host particles located at $X$. 

Recall that \autoref{eq:dE_df} gives the total energy, $dE\sub{DF}(r)$, transferred from the substructure to the host through DF, when the substructure passes through the radius, $r$. We suppose that this corresponds to the integration of $dN \times dK$ over $b$ and $\varphi$,
\begin{equation}
    dE\sub{DF}(r) = A \iint \frac{\rho(D)b}{[1 + (b/b\sub{90})^2]^2} d \varphi db \equiv A \iint W d \varphi db,
        \label{eq:redistribution}
\end{equation}
where $A$ is a constant. Note that $D = D(b,\varphi)$ for given ${\bf r}$ and ${\bf v}$ and thus the kernel function, $W = W(b, \varphi)$. Since the mass of each host particle is much smaller than that of the substructure, we compute $b\sub{90} = GM/v^2$ with the velocity of the substructure, $v$ (\autoref{eq:v}). Integrating \autoref{eq:redistribution} over the range of $\varphi = [0 : 2\pi]$ and $b = [0 : r'+R']$, the constant, $A$, is derived and the energy redistribution to a given radius is determined based on the kernel function, $W$.

\subsection{Tidal shock}
\label{ssec:tidal_shock}

When a substructure passes by an extended host system, the host is tidally perturbed. Because gravity is a long range force, this process can cause global effects onto the host. The sensitivity of host particles to the tidal perturbation depends on the orbital period of each host particle and the timescale of the variation of the gravitational potential caused by the substructure. If the former is shorter than the latter, the host particle reacts to the potential variation adiabatically and no change is made by the potential variation, i.e., adiabatic shielding \citep{Spitzer1987,Gnedin1999_adiabatic_shield}. On the other hand, the motion of host particles having orbital periods longer than the timescale of the tidal interaction is changed by the tidal force and can gain kinetic energy, the so-called tidal shock \citep[TS; e.g.,][]{Spitzer1958,Aguilar1985,Gnedin1999_tidal_shock}.

We study how much of energy is deposited into the host by TS. For simplicity, we suppose that the substructure orbiting in the host system is a point mass, $M$. The pericentre, $r\sub{p}$, is the place of the shortest timescale of the potential variation caused by the substructure. For most of host particles, the use of the impulse approximation in which the position of the host particle is unchanged during the potential variation is justified at $r\sub{p}$. According to \cite{Spitzer1958}, the specific energy that a host particle located at ${\bf r}$ gains in the TS process is
\begin{equation}
    d \epsilon\sub{TS}(r) = \frac{4}{3} \biggl ( \frac{G M r}{r\sub{p} l} \biggr )^2.
        \label{eq:depsts}
\end{equation}
The model assumed straight-line orbits for simplicity, while \autoref{eq:depsts} shows good agreements with the updated model for eccentric orbits \citep{Gnedin1999_tidal_shock,Banik2021}. Note that adiabatic shielding is not taken into account in \autoref{eq:depsts}, and the actual efficiency of energy gain would be lower at small radii where the orbital period of host particles becomes short. Despite neglecting adiabatic shielding, $d \epsilon\sub{TS}$ in \autoref{eq:depsts} decays as $r$ approaches zero. Comparing \autoref{eq:depsts} to the host potential, $\Phi(r)$, we know the significance of the energy deposit to the host due to TS. Since host particles gain kinetic energy through TS, the substructure loses its orbital energy as a back-reaction. The amount of the lost energy from the substructure at $r$ compensates the energy gained by the mass shell at $r$,
\begin{equation}
    dE\sub{TS}(r) = 4 \pi r^2 \rho(r) dr d \epsilon\sub{TS}(r).
        \label{eq:dets}
\end{equation}

\subsection{Orbital and mass evolution of the substructure}
\label{ssec:substructure_evo}

As DF and TS reduce the orbital energy and angular momentum of the substructure, its orbit decays with time. For updating the orbital energy and angular momentum of the substructure, \autoref{eq:dE_df}, \autoref{eq:dL}, and \autoref{eq:dets} are used. The loss of energy and angular momentum is integrated over the orbital period and those quantities are updated at the apocentre. The mass of the substructure is also reduced by the tidal force of the host system (tidal stripping). At each pericentre, $r\sub{p}$, we measure the tidal radius of the substructure,
\begin{equation}
    r\sub{t} = r\sub{p} \biggl [ \frac{M\sub{s}(r\sub{t})}{2M\sub{h}(r\sub{p})} \biggr ]^{1/3} , 
\end{equation}
where $M\sub{s}(r)$ and $M\sub{h}(r)$ are the mass profiles of the substructure and the host, respectively. Here, the both systems are considered as extended objects initially following the NFW density profile, while the substructure is treated as a point mass in the models of DF and TS. The mass of the substructure outside $r\sub{t}$ is assumed to be stripped off from the substructure, and its density structure is truncated at $r\sub{t}$. This treatment is justified by numerical simulations \citep[e.g.,][]{Penarrubia2010,Drakos2020}.

\subsection{Dynamical heating by DF and TS}
\label{ssec:dyn_heating}

\begin{figure}
    \begin{center}
        \includegraphics[width=0.4\textwidth]{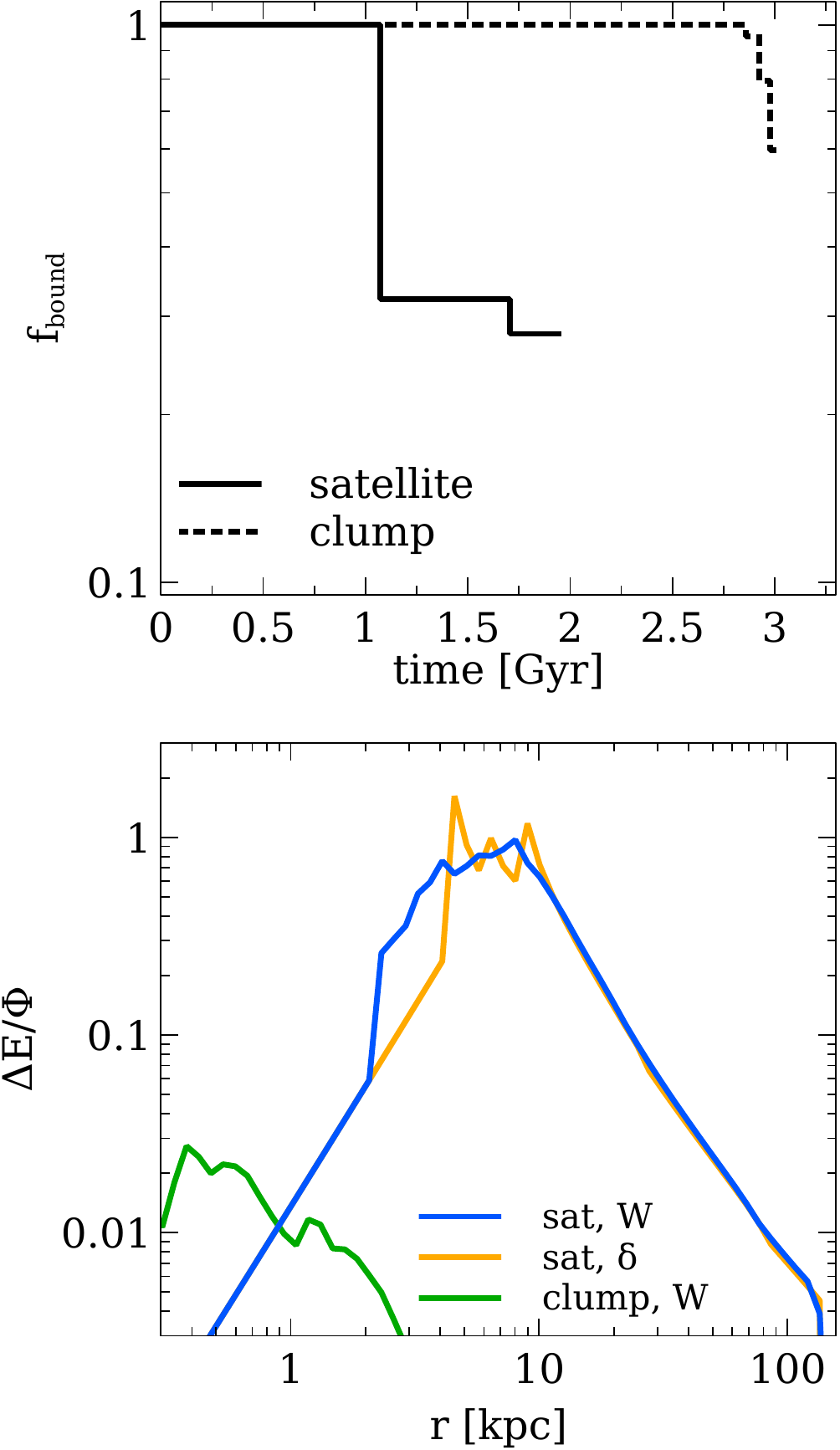}
    \end{center}
        \caption{
            Prediction by the analytical model. 
            {\it Upper panel:} Mass evolution of the satellite (solid) and clump (dotted). 
            {\it Lower panel:} Ratio of the specific energy deposition to the local potential of the host. Blue and orange lines study the dynamical heating by the satellite with the $W$- and $\delta$-kernel functions, respectively. Green line shows the dynamical heating by a single giant clump, employing the $W$-kernel. The dynamical heating of the satellite can alter the dynamical structure of the host around the pericentre ($3 \la r/{\rm kpc} \la 10$). Considering multiple clumps, they would be a promising heating source to flatten the cusp.
        \label{fig:ana_m_heat}}
\end{figure}

The dynamical heating by a substructure onto the host is discussed using the analytical model. We consider two heating sources, a satellite galaxy and a single baryonic clump, corresponding to the merger and clump simulations, respectively. The structure of the host system is described in \autoref{sec:sim_basics}. In the analysis of the galaxy merger model, the structural parameters of the satellite and the merger orbit are the same as employed in the merger simulation (\autoref{ssec:merger_setup}). While clump simulations include multiple clumps, only a single clump is considered in the analysis of the clump heating. The structure of the clump is as employed in the run-A (\autoref{ssec:clump_setup}), and the clump is initialised on a circular orbit of $r=10$~kpc. We adopt the Coulomb logarithm of $\ln{\Lambda}=2.4$ in the analyses \citep[e.g.,][]{Taylor2001}. 

The upper panel of \autoref{fig:ana_m_heat} shows the predicted mass evolution of the satellite (solid) and clump (dotted). The satellite experiences the dynamical evolution of three pericentric passages by $t=2.0$~Gyr and merges with the host due to the loss of the orbital energy and angular momentum. At the time of merging, $\sim$ 30 percent of its mass is retained, consistent with the \nbody simulation result (lower panel of \autoref{fig:d21_orb_mass}). The tidal massloss of the substructure is considered at each pericentric passage, and the top panel also indicates that the orbital period gets shorter with time because of the orbital decay driven by DF and TS. The clump keeps its mass perfectly at $t \la 3$~Gyr because of the high density. As its orbit decays with time, its mass decreases in the last a few orbits before sinking to the host centre.

In the lower panel of \autoref{fig:ana_m_heat}, we compare the specific energy deposited into the host by DF and TS, $\Delta E \equiv d \epsilon\sub{DF} + d \epsilon\sub{TS}$, to the gravitational potential of the host, $\Phi(r)$. We first study the galaxy merger scenario. The analysis of the blue line employs the $W$-kernel function for modelling the global DF dynamical heating (\autoref{sssec:df_global}). In the the analysis of the orange line, DF is considered as a local process (\autoref{sssec:df_local}), and the corresponding kernel function is Dirac's $\delta$-function. We find that the deposited energy at $3 \la r/{\rm kpc} \la 10$ is comparable to $\Phi(r)$, and thus the dynamical heating by the satellite can alter the density structure of the host within this radial range. Even when considering DF as a global process (blue), the energy deposition is locally limited and the heating in the central kpc of the host would be insufficient to flatten the cusp. The green line shows that the clump heats up the host centre more efficiently than the satellite. Considering multiple clumps, the central cusp may be flatted out. 

\begin{figure}
    \begin{center}
        \includegraphics[width=0.4\textwidth]{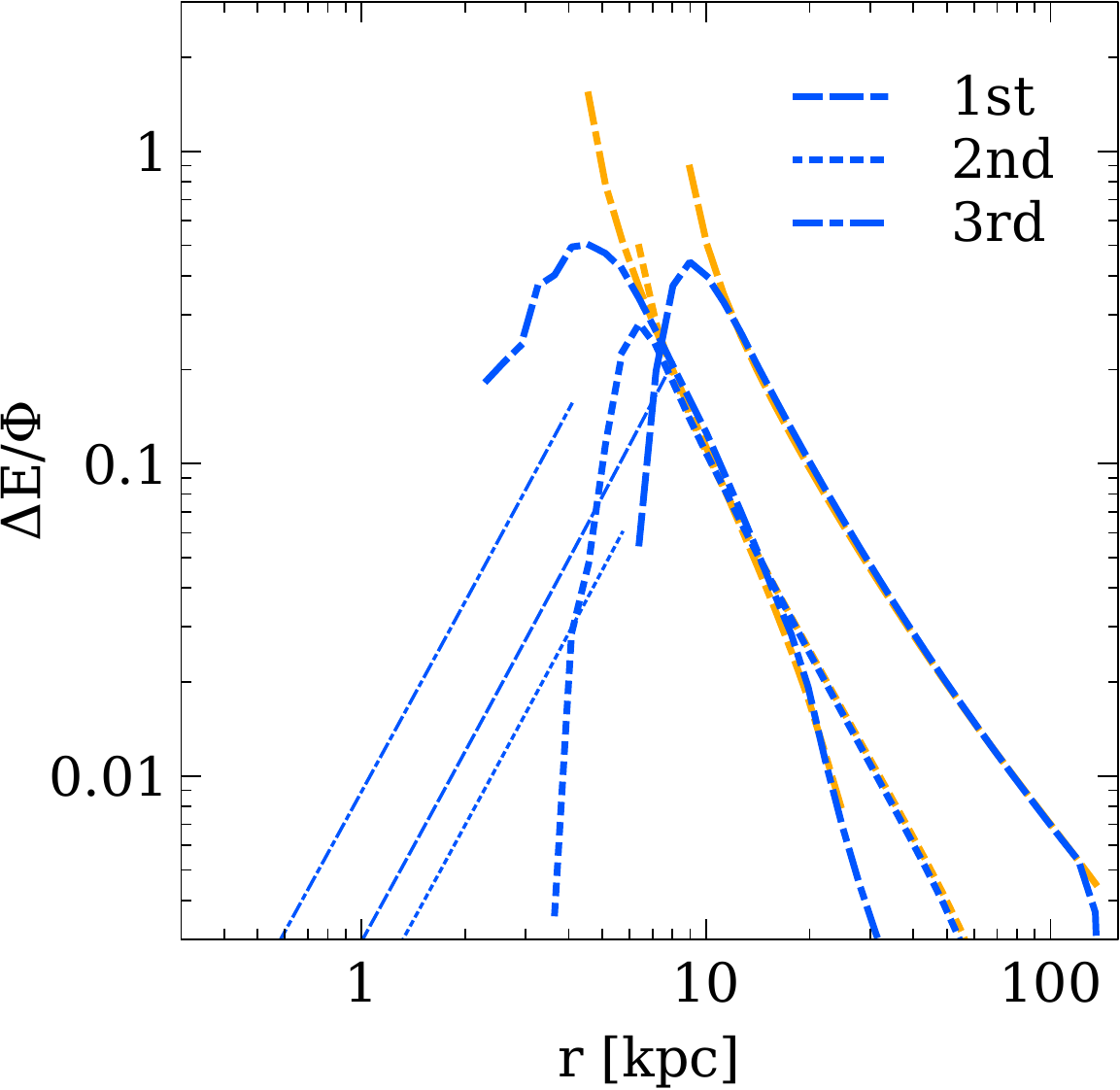}
    \end{center}
        \caption{
            Decomposition of the dynamical heating by the satellite. As in \autoref{fig:ana_m_heat}, blue and orange lines employ the $W$- and $\delta$-kernel functions, respectively. Dashed, dotted and dot-dashed line styles represent the dynamical heating in the first, second and third orbital periods. Thick and thin lines indicate the contributions from DF and TS, respectively. Even if considering DF as a global process (blue; see \autoref{sssec:df_global}), it is difficult to heat up the central kpc of the host.
        \label{fig:ana_heat_breakdown}}
\end{figure}

To understand the result presented in the lower panel of \autoref{fig:ana_m_heat}, we decompose the energy deposition by the satellite into the contribution of the first (dashed), second (dotted) and third orbital periods (dot-dashed) in \autoref{fig:ana_heat_breakdown}. The heating by DF (TS) is represented by thick (thin) lines. When employing the $\delta$-kernel (orange), the energy deposition by DF sharply increases at the pericentre, while no-energy is deposited to the inner radii. This feature makes the peaks found in the lower panel of \autoref{fig:ana_m_heat} ($r = 4$, 7 and 10~kpc). In the model employing the $W$-kernel (blue), the energy is deposited into the radial range inside the pericentre, while the energy deposition profile peaks at each pericentre. The extension of the energy deposition by the $W$-kernel is limited. Even when employing the $W$-kernel, the heating by DF does not reach the central kpc. 

These results indicate that substructures need to approach the centre of the host to heat up the cusp in the NFW halo, despite the fact that gravity is a long range force. These results are consistent with the findings by \cite{Dekel2021}, who showed that when the satellite is diffuse, it is disrupted by the tidal force of the host before approaching to the host centre and the heating efficiency of the cusp is significantly lowered. On the other hand, the tidally resilient satellite can survive in the strong tidal field and heat up the cusp at the centre of the host. Also note that the deposited energy can be re-distributed to the host centre through violent relaxation \citep{Lynden-Bell1967}, even if DF or TS does not directly heat up the host centre.